\documentclass[prb,onecolumn,superscriptaddress,showpacs,10pt]{revtex4-2}

\usepackage{bm}
\usepackage{graphicx}
\usepackage{amsmath}
\usepackage{hyperref}
\usepackage{epsfig}

\usepackage{epstopdf}

\usepackage[ngerman, english]{babel}

\DeclareGraphicsExtensions{.eps}

\usepackage{hyperref}
\hypersetup{%
	colorlinks=true,
	linkcolor=blue,
	urlcolor=blue,
	citecolor=blue
}

\begin{document}
	
	\title{ Effect of impurity scattering on percolation of bosonic islands and reentrant superconductivity in Fe implanted NbN thin films}

	\author{Rajdeep Adhikari}
	\email{rajdeep.adhikari@jku.at}
	\affiliation{Institut f\"ur Halbleiter-und-Festk\"orperphysik, Johannes Kepler University, Altenbergerstr. 69, A-4040 Linz, Austria}

	\author{Bogdan Faina}
	\affiliation{Institut f\"ur Halbleiter-und-Festk\"orperphysik, Johannes Kepler University, Altenbergerstr. 69, A-4040 Linz, Austria}
	
	\author{Verena Ney}
	\affiliation{Institut f\"ur Halbleiter-und-Festk\"orperphysik, Johannes Kepler University, Altenbergerstr. 69, A-4040 Linz, Austria}

	\author{Julia Vorhauer}
	\affiliation{Institut f\"ur Halbleiter-und-Festk\"orperphysik, Johannes Kepler University, Altenbergerstr. 69, A-4040 Linz, Austria}

	\author{Antonia Sterrer}
	\affiliation{Institut f\"ur Halbleiter-und-Festk\"orperphysik, Johannes Kepler University, Altenbergerstr. 69, A-4040 Linz, Austria}

	\author{Andreas Ney}
	\affiliation{Institut f\"ur Halbleiter-und-Festk\"orperphysik, Johannes Kepler University, Altenbergerstr. 69, A-4040 Linz, Austria}

	\author{A. Bonanni}
	\email{alberta.bonanni@jku.at}
	\affiliation{Institut f\"ur Halbleiter-und-Festk\"orperphysik, Johannes Kepler University, Altenbergerstr. 69, A-4040 Linz, Austria}

	\begin{abstract}
		A reentrant temperature dependence of the thermoresistivity $\rho_{\mathrm{xx}}(T)$ between an onset local superconducting ordering temperature $T_\mathrm{loc}^\mathrm{onset}$ and a global superconducting transition at $T=T_\mathrm{glo}^\mathrm{offset}$ has been reported in disordered conventional 3-dimensional (3D) superconductors. The disorder of these superconductors is a result of either an extrinsic granularity due to grain boundaries, or of an intrinsic granularity ascribable to the electronic disorder originating from impurity dopants. Here, the effects of Fe doping on the electronic properties of sputtered NbN layers with a nominal thickness of 100 nm are studied by means of low-$T$/high-$\mu_{0}H$ magnetotransport measurements. The doping of NbN is achieved $via$ implantation of 35 keV Fe ions. In the as-grown NbN films, a local onset of superconductivity at $T_\mathrm{loc}^\mathrm{onset}=15.72\,\mathrm{K}$ is found, while the global superconducting ordering is achieved at $T_\mathrm{glo}^\mathrm{offset}=15.05\,\mathrm{K}$, with a normal state resistivity $\rho_{\mathrm{xx}}=22\,{\mu\Omega}\cdot{\mathrm{cm}}$. Moreover, upon Fe doping of NbN, $\rho_{\mathrm{xx}}=40\,{\mu\Omega}\cdot{\mathrm{cm}}$ is estimated, while $T_\mathrm{loc}^\mathrm{onset}$ and $T_\mathrm{glo}^\mathrm{offset}$ are measured to be 15.1 K and 13.5K, respectively. In Fe:NbN, the intrinsic granularity leads to the emergence of a bosonic insulator state and the normal-metal-to-superconductor transition is accompanied by six different electronic phases characterized by a $N$-shaped $T$ dependence of $\rho_{\mathrm{xx}}(T)$. The bosonic insulator state in a $s$-wave conventional superconductor doped with dilute paramagnetic impurities is predicted to represent a workbench for emergent phenomena, such as gapless superconductivity, triplet Cooper pairings and topological odd frequency superconductivity. 
	\end{abstract}
	
	\date{\today}
	
	
	\maketitle

	\section{Introduction}
	
	The doping of superconductors with non-magnetic and magnetic impurities paved the way for understanding the physics of isotropic and anisotropic superconductivity \cite{Balatsky:2006_RMP,Mockli:2020_JAP,Gorkov:2008_Book}. The effects of impurities on the electronic properties of both conventional and unconventional superconductors were explained in terms of electron lifetime and pairing symmetries due to scattering by an ensemble of impurities \cite{Anderson:1959_JPCS,Abrikosov:1961_JETP,Abrikosov:1969_JETP,Balatsky:2006_RMP,Zittartz:1970_ZPhys_1,Mueller-Hartmann:1970_ZPhys,Zittartz:1970_ZPhys_2,Woolf:1965_PhysRev,Skalski:1964_PhysRev,Fulde:1966_PhysRev,Fulde:1973_Adv.Phys.,Maki:1967_PhysRev,Anderson:1959_JPCS}. According to the Cooper's pairing model \cite{Cooper:1956_PhysRev} and to the Bardeen-Cooper-Schreiffer's (BCS) theory, \cite{BCS:1957_PhysRev_1,BCS:1957_PhysRev_2}, superconductivity is a result of the instability of the Fermi surface against the pairing of time-reversed quasiparticle states. Any perturbation that is unable to lift the Kramers degeneracy of these states does not affect the superconducting transition temperature. While non-magnetic or scalar impurities are not known to affect the isotropic singlet $s$-wave order parameter in conventional BCS superconductors, magnetic impurities efficiently break the Cooper pairs, leading to the suppression of any long range superconducting order \cite{Balatsky:2006_RMP} or to the emergence of gapless superconductivity \cite{Woolf:1965_PhysRev,Abrikosov:1969_JETP,Balatsky:2006_RMP}. Most theoretical approaches \cite{Gorkov:2008_Book,Fulde:1966_PhysRev,Pint:1989_PhysC} based on the BCS theory \cite{BCS:1957_PhysRev_1,BCS:1957_PhysRev_2} treat the interaction of the impurity spin $\vec{S}$ with the spin $\vec{\sigma}$ of the conduction electrons $via$ an exchage interaction $\vec{\sigma}\cdot{\vec{S}}$. For a magnetic impurity embedded in the superconducting matrix, there is a coupling between the local spin on the impurity site and the conduction electrons of the matrix represented by the Hamiltonian:
	
	\begin{equation}
		H_\mathrm{imp}=\sum_{\alpha\beta}^{}\int_{}^{}d\vec{r} J(\vec{r})\Psi^{\dagger}_{\alpha}(\vec{r})\hat{U}^\dagger_{\mathrm
			imp}\Psi_{\beta}(\vec{r})
	\end{equation}
	\noindent
	
	where $\hat{U}^\dagger_{\mathrm{imp}}=J(\vec{r})\vec{\sigma}\cdot\vec{S}$ \cite{Maki:1967_PhysRev} and $J(\vec{r})$ is the exchange interaction between the impurity ion and the conduction electrons while $\alpha$ and $\beta$ represent the particle indices. Furthermore, it was also predicted that single magnetic impurities lead to pair breaking and result in the formation of quasiparticle Yu-Shiba-Rusinov (YSR) states within the energy gap and localized in the vicinity of the impurity atom \cite{Heinrich:2018_PSS}. However, for dilute magnetic impurities with atomic concentrations $\lesssim{10^{-4}}\%$ the spins can be considered randomly oriented and uncorrelated \cite{Pint:1989_PhysC}.
	Another approach to study the effect of magnetic impurities on superconductivity is to map the nature of superconductivity with a ferromagnetic background resulting from the proximity of a superconducting thin film to magnetic layers, promoting the onset of emergent phenomena, including spin-triplet Cooper pairing, Majorana fermions and spin superfluids \cite{Linder:2015_NatPhys,Mueller:2021_PRL}. The nature of superconductivity in strong spin-exchange fields was extensively discussed by Fulde-Ferrel \cite{Fulde:1964_PhysRev} and Larkin-Ovchinnikov \cite{Larkin:1964_JETP}. The existence of the Fulde-Ferrel-Larkin-Ovchinnnikov (FFLO) pairing was reported for heavy fermion systems, organic superconductors and recently for combinations of superconductors (S) and ferromagnets (F) in F/S/F and S/F/S heterostructures \cite{Lenk:2016_PRB,Buzdin:2005_RMP}.

	Within the family of the conventional superconductors, NbN with a bulk superconducting transition temperature of $\sim$16 K, has been widely studied both in the bulk crystal phase and as thin film \cite{Chockalingam:2008_PRB,Koushik:2013_PRL,Ganguly:2015_PRB,Chand:2009_PRB,Chand:2012_PRB,Mondal:2011_PRL,Chockalingam:2009_PRB,Destraz:2017_PRB}. Magnetron sputtered NbN thin films are intensively investigated \cite{Mondal:2011_PRL,Chand:2009_PRB,Chand:2012_PRB,Chockalingam:2009_PRB,Ganguly:2015_PRB} in view of diverse relevant applications of superconducting NbN in Josephson junctions, hot electron bolometers, single photon detectors \cite{Nikzad:2016_Sensors,Polakovic:2020_Nanomater} and in devices for quantum information and circuit quantum electrodynamics \cite{Blais:2021_RMP}. Disordered NbN thin films grown by sputtering are used as the workbench to investigate the Berezinskii-Kosterlitz-Thouless (BKT) phase transition \cite{Mondal:2011_PRL,Yong:2013_PRB}, superconductor-insulator transitions \cite{Mondal:2011_JSNM,Chand:2012_PRB}, conductance fluctuation \cite{Koushik:2013_PRL}, Andreev reflection and the Higgs-Anderson mechanism of superconductivity \cite{Sherman:2015_NatPhys,Tsuji:2020_PRRes}. Doping of conventional superconductors like NbN with magnetic impurities is expected to broaden the perspectives for hybrid structures-based applications in superconducting spintronics \cite{Linder:2015_NatPhys}, spin-orbitronics, in dark matter detectors \cite{Hochberg:2016_PRL}, integrated resonators and superconducting qubit processors for quantum computation.

	While superconducting layers of NbN are traditionally grown on conventional substrates such as Si and MgO, the use of GaN and Al$_{1-x}$Ga$_{x}$N templates and substrates has emerged as an alternative \cite{Krause:2014_SST,Sam-Giao:2014_AIPAdv,Kobayashi:2020_APE,Kobayashi:2020_APL} and can serve as basis for all nitride integrated superconductor/semiconductor devices \cite{Yan:2018_Nature}. The Si substrates are commonly employed to grow superconducting NbN films intended for single photon detection and hot bolometer applications, due to the advantages in device processing and relatively low losses at THz frequencies. The significant lattice mismatch between Si and NbN generally results in polycrystalline films with $T_\mathrm{c}\le{10}$~K for layers $\sim(5-10)$ nm thick. The lattice matched substrate MgO presents challenges in device processing, due to its hydrophobic nature and sensitivity to alkaline solutions that are used during the fabrication processes \cite{Krause:2014_SST}. Since the crystallographic orientation of the NbN films does not affect their superconducting properties, the $c$-plane of hexagonal wurtzite template/buffer layers is also suitable for NbN growth, provided that the lattice parameter $a^{\mathrm{w}}_\mathrm{hkl}$ of the wurtzite template matches the one of NbN along the (111) plane $i.e.$ $a^{\mathrm{NbN}}_{111}=\frac{a^{\mathrm{w}}_\mathrm{100}}{\sqrt{2}}$. Thus, the epitaxial relation $\left [ 111 \right ] \left ( \mathrm{NbN} \right )\parallel \left [ 100\right ] \left ( \mathrm{GaN} \right )$ is established between NbN and $c$-plane wurtzite template\cite{Krause:2014_SST,Kobayashi:2020_APL,Yan:2018_Nature}.

	The effect of dilute magnetic impurities on the superconducting properties of NbN thin films is scarcely reported in literature. The ferromagnetic proximity in the NbN/FeN bilayer system has been shown to lower the superconducting transition temperature \cite{Hwang:2017_PhysC}, and Gd ion implantation was reported to have similar effects on the superconducting properties of sputtered NbN films \cite{Jha:2013_JSNM}. In the following, the emergence of a bosonic insulator (BI) state in Fe implanted sputtered NbN thin film using low-$T$/high-$\mu_{0}H$ magnetotransport studies is analysed.

	
	\section{Samples and Experimental Details}
	
	\begin{figure*}[htbp]
		\centering
		\includegraphics[scale=1.25]{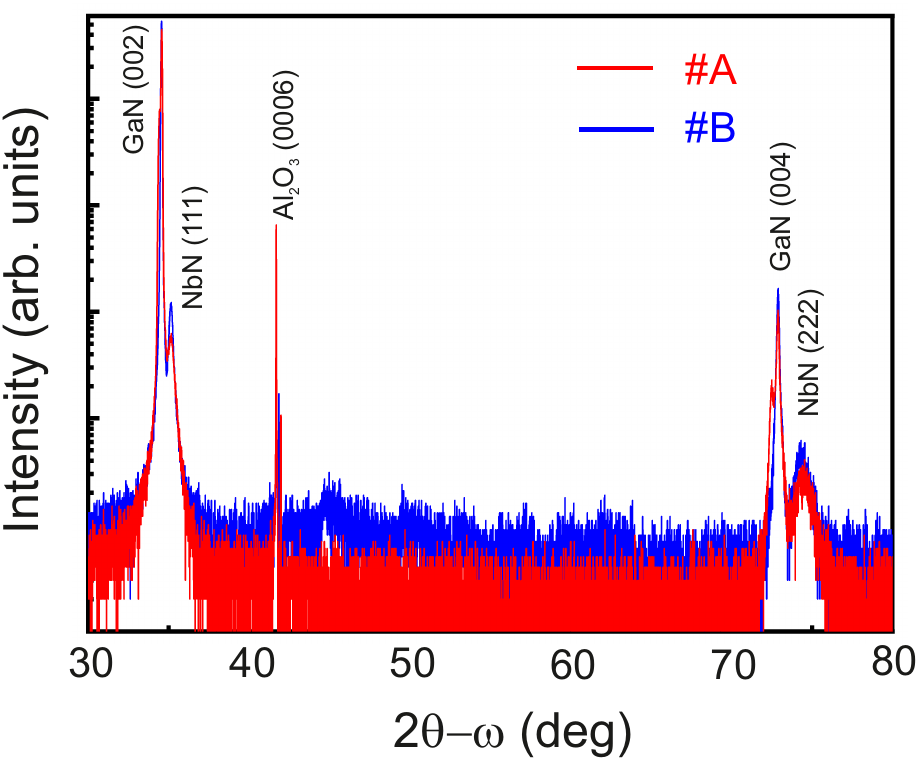}
		\caption{\label{fig:XRD} High resolution ($2{\theta}-{\omega}$) scan of \#A and \#B.}	
		\label{fig:fig1}
	\end{figure*}
	
	\subsection{Growth and structure}

	The NbN films considered in this work are grown by reactive magnetron sputtering with high purity Nb as target. Epitaxial single crystalline $\sim$ 1 µm GaN grown on one-side polished $c$-plane Al$_{2}$O$_{3}$ (0001) substrates by means of metal organic vapor phase epitaxy is taken as template for the deposition of the NbN thin films. The GaN templates are cut into $\left(1\times1\right)$ cm$^{2}$ specimens for the sputtering of the NbN layers. The polycrystalline thin films are grown in an ultrahigh vacuum (UHV) chamber with a base pressure of $(2\times{10^{-9}})$ mbar. A high purity (99.99\%) Nb target is employed for the reactive magnetron sputtering process under a plasma consisting of Ar:N$_{2}$ in the ratio 10:5 standard cubic centimeters per minute (sccm) with a power $P=40$ W and at a constant substrate temperature $T_\mathrm{sub}=500^{\circ}$C. 
	
	\begin{table*}
		
		\centering
		
		\caption{Relevant parameters of the representative samples considered in this work.}
		
		\renewcommand{\arraystretch}{2.0}
		
		\resizebox{\columnwidth}{!}{\begin{tabular}{|l|l|l|l|l|l|l|l|l|l|}
				\hline
				Sample & Material & Template & Nominal $d_\mathrm{NbN}$ (nm) &  Ar:N$_{2}$ ratio  & P (W) & T$_\mathrm{sub}$ (°C) & Implanted ion & $E_\mathrm{ion}$ (keV) & Dose (at/cm$^{3}$) \\ \hline
				A          & NbN      & wz-GaN   & 100    & 10:5 & 40    & 500       & -                 & -                  & -            \\ \hline
				B          & Fe:NbN   & wz-GaN   & 100    & 10:5 & 40    & 500       & Fe                & 35                 & $1 \times 10^{14}$    \\ \hline
		\end{tabular}}
		\label{tab:tab1}
	\end{table*}

	The NbN films are implanted with $^{56}{\mathrm{Fe}}$ using powdered Fe$_{3}$O$_{4}$ as the source for Fe ions. The implantation is carried out at room temperature at a base pressure of $(1\times{10^{-6}})$ mbar. Half of the surface of the $\left(1\times1\right)$ cm$^{2}$ samples is covered with a brass mask and the uncovered part of the NbN layers is exposed to the Fe ion beam for implantation. The energy and ion dose are selected after simulating the stoppage and range of ions in matter (SRIM). The simulation of the ion distribution is considered for ion energies of 35 keV and 50 keV, respectively. A low ion current $\sim$ 100 nA is required to ignite the plasma and for the subsequent ion extraction. The ion current is kept constant by tuning the filament current and ion beam focus during the entire implantation process. The Fe implantation dose used in this work is kept constant at $\left(1\times10^{14}\right)$ atoms/cm$^{2}$. After implantation, the samples are not subjected to any thermal annealing, in order to avoid precipitation of secondary Fe or FeN phases and also to preserve the defects generated by the implanted ions. The $\left(1\times1\right)$ cm$^{2}$ samples are then cut into $\left(5\times5\right)$ mm$^{2}$ specimens for further measurements and characterization. In the following, a representative as-grown NbN layer sample A (\#A), is considered, while a 35 keV Fe implanted NbN layer is taken as sample B (\#B). Both the as-grown NbN and Fe implanted NbN are studied using x-ray diffraction (XRD) and the high crystallinity of the sputtered films is confirmed \cite{Vorhauer:2021_Thesis}. An overview of the relevant growth parameters for \#A and \#B is provided in Table I. 
	
	Radial scans are acquired using a PANalytical X'Pert PRO Materials Research Diffractometer (MRD) equipped with a hybrid monochromator with a  divergence slit of 1/4$^\circ$. The diffracted beam is collected by a solid-state PixCel detector provided with a 9.1 mm anti-scatter slit. The $\left(2\theta-\omega\right)$ scans for \#A and \#B are shown in  Fig.~\ref{fig:fig1}. The Bragg peaks of NbN(111) and NbN(222) are observed alongside the GaN(002) and GaN(004) peaks, as indicated in Fig.~\ref{fig:fig1}. The obtained result also points at an epitaxial relation $a^{\mathrm{NbN}}_{111}=\frac{a^{\mathrm{w}}_\mathrm{100}}{\sqrt{2}}$ of the NbN (111) on $c$-Al$_2$O$_3$ (0006) as reported in literature \cite{Krause:2014_SST}. No secondary Bragg peaks of precipitated or clustered Fe and Fe$_{4}$N are detected, pointing out the homogeneous incorporation of the implanted Fe in the cubic NbN crystal lattice. 
	
	\subsection{Magnetotransport}

	The low-$T$/high-$\mu_{0}H$ magnetotransport studies are performed on \#A and \#B in a four probe van der Pauw geometry. Ohmic contacts to the $(5\times{5})$ mm$^{2}$ specimens are achieved using electrically conducting Ag epoxy and are bonded with Au wires of diameter 25 $\mu$m. The magnetotransport experiments are carried out in a Janis Super Variable Temperature 7TM-SVM cryostat (Janis Cryogenics, Westerville, OH, USA) equipped with a 7 T superconducting magnet. A lock-in amplifier (LIA) ac technique is employed for measuring the magnetotransport properties of the NbN and of the Fe:NbN thin films. The constant current $I_\mathrm{ac}$ is sourced from a Stanford Research SR830 LIA $via$ a resistance decade box, while the longitudinal voltage $V_\mathrm{xx}$ is measured in a phase-locked mode. The lock-in expand function is employed to enhance the sensitivity of the LIA. All measurements have been performed at a frequency of 127 Hz, while $I_\mathrm{ac}=1$~$\mu$A for all measurements. The low input current minimizes the thermal drift due to Joule heating of the samples. The magnetic fields are varied between $-7$ T and $+7$ T for the magnetoresistance measurements. 
	
	\begin{figure*}[htbp]
		\centering
		\includegraphics[scale=2.50]{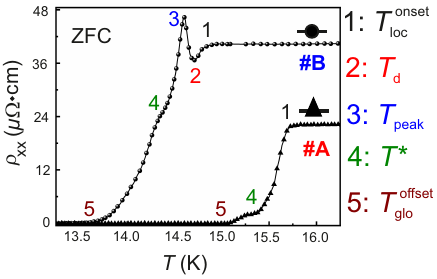}
		\caption{\label{fig:ZFC} ZFC $\rho_{xx}$ as a function of $T$ for samples \#A and \#B.}	
		\label{fig:fig2}
	\end{figure*}
	
	All experimental set-ups including the LIA, the magnet power supply, and the temperature controller	are regulated by means of an indigenously developed software. Prior to the measurements, the Ohmic characteristics of the Ag epoxy contacts to the samples are tested with a high resolution Keithley 4200 SCS source-measure unit (SMU). The longitudinal resistivity $\rho_\mathrm{xx}$ is estimated as a function of temperature $T$ and magnetic field $\mu_{0}H$ from the measured longitudinal voltage $V_\mathrm{xx}$. The evolution of thermoresistivity $\rho_{\mathrm{xx}}(T)$ as a function of $T$ is estimated by cooling the sample both in the absence and in the presence of a transverse magnetic field $\mu_{0}H_\perp$ and of a longitudinal magnetic field $\mu_{0}H_\parallel$. While $\mu_{0}H_\perp$ is applied parallel to the surface normal of the sample, $\mu_{0}H_\parallel$ is applied perpendicular to it. The zero field cooled (ZFC) $\rho_\mathrm{xx}-T$ behavior is studied while the samples are cooled down for $\mu_{0}H_{\perp}=0$ ($\mu_{0}H_{\parallel}=0$) and the field cooled (FC) for $\mu_{0}H_{\perp}\ne0$ ($\mu_{0}H_{\parallel}\ne0$).

	\section{Results} 
	
	The evolution of the ZFC $\rho_\mathrm{xx}$, measured as a function of $T$ in the interval $10~\mathrm{K}\le{T}\le16.5~\mathrm{K}$ for samples \#A and \#B is reported in Fig.~\ref{fig:fig2}. The normal state resistivity $\rho_{\mathrm{xx}}$ of \#B at $T\ge{T_\mathrm{loc}^\mathrm{onset}}$ is recorded to be $\sim$40 $\Omega\cdot{\mathrm{cm}}$, while for \#A, $\rho_{\mathrm{xx}}$ $\sim$22 $\Omega\cdot{\mathrm{cm}}$. The presence of Fe dopants in the NbN matrix is identified to be the reason for the increase of $\rho_{\mathrm{xx}}$. As evidenced in Fig.~\ref{fig:fig2}, the ZFC $\rho_\mathrm{xx}(T)$ of \#B presents five characteristic temperatures labeled by 1, 2, 3, 4 and 5. The characteristic temperatures 1, 4 and 5 are found also for the ZFC $\rho_\mathrm{xx}(T)$ of \#A. These characteristic temperatures are marked as follows:	1: $T_\mathrm{loc}^\mathrm{onset}$;	2: $T_\mathrm{d}$;	3: $T_\mathrm{peak}$; 4: $T = T^{*}$ and 5: $T_\mathrm{glo}^\mathrm{offset}$.

	It is also evidenced in  Fig.~\ref{fig:fig2}, that the thermodynamic transition from the normal state N to the superconducting state S occurs over a temperature range $\Delta{T_\mathrm{c}}$ defined as: 
	\begin{equation}
		\Delta{T_\mathrm{c}}=\left(T_\mathrm{loc}^\mathrm{onset}-T_\mathrm{glo}^\mathrm{offset}\right)
	\end{equation}
	\noindent
	
	\begin{figure*}[htbp]
		\centering
		\includegraphics[scale=1.25]{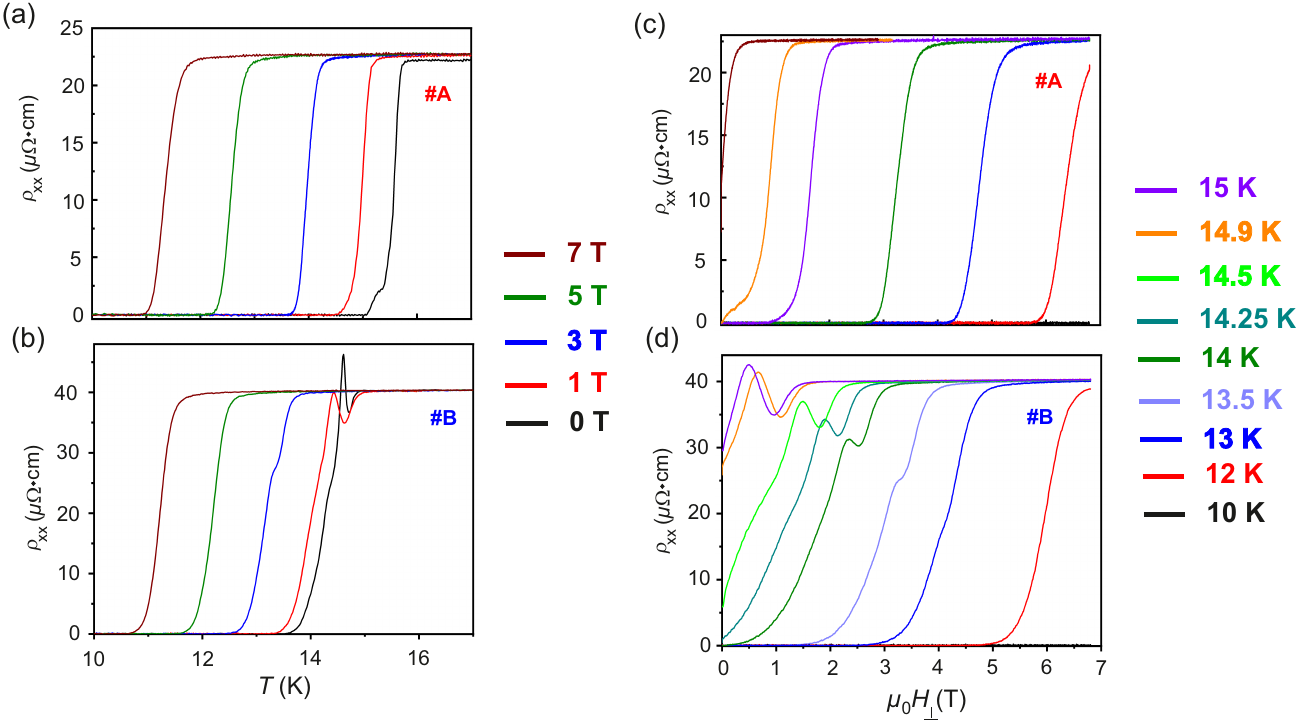}
		\caption{\label{fig:RTMR}  ZFC and FC $\rho_{xx}$ as a function of $T$ under the application of $\mu_{0}H_\perp$ = 0 T, 1 T, 3 T, 5 T and 7 T measured for samples (a) \#A and (b) \#B. Transverse $\rho_\mathrm{xx}$ as a function of $\mu_{0}H_\perp$ recorded at different $T$ for samples (c) \#A and (d) \#B.}	
		\label{fig:fig3}
	\end{figure*}

	In the as-grown NbN film \#A, the dominant microstructural disorder originates from the grain boundaries of the polycrystalline film, as reported for similar polycrystalline and amorphous systems \cite{Zhang:2013_PRL,Zhang:2016_PRApp,Postolova:2017_SciRep,Mondal:2011_JSNM,Ganguly:2015_PRB}. Such a 3-dimensional (3D) system with only a structural disorder is expected to undergo a N/S transition with delayed onset of the global superconductivity. These disordered superconductors are characterized by a finite non-zero value of $\Delta{T_\mathrm{c}}$ \cite{Zhang:2016_PRApp}. The physical mechanism of this transition is described in terms of the percolation of  bosonic clusters - composed of Cooper pairs - to form bosonic islands at the $T=T_\mathrm{loc}^\mathrm{onset}$ indicated in Fig.~\ref{fig:fig2}. These bosonic islands percolate to produce a network of bosonic conduction channels, which in turn replaces the single particle fermionic conduction channels of the N state of NbN. The grain boundaries serve as scattering centers for the bosonic islands, leading to the observed kink at $T=T^{*}$, as indicated in the ZFC $\rho_{\mathrm{xx}}(T)$ curve in Fig.~\ref{fig:fig2}. At $T=T_\mathrm{glo}^\mathrm{offset}$, the global superconducting phase sets in, as evidenced in Fig.~\ref{fig:fig2}.

	However, in \#B two more characteristic temperatures indicated by $T_\mathrm{d}$ and $T_\mathrm{peak}$ are observed in the evolution of $\rho_{\mathrm{xx}}(T)$, as presented in Fig.~\ref{fig:fig2}. The anomalous peak at $T=T_\mathrm{peak}$ in the $\rho_{\mathrm{xx}}(T)$ points at the existence of disorder induced electronic phase transitions for the N/S thermodynamic phase transition in \#B, due to the presence of the implanted Fe ions. In order to understand the mechanism of the electronic processes involved in the observed behavior of \#B, FC $\rho_{\mathrm{xx}}(T)$ measurements are performed on both \#A and \#B for $\mu_{0}H_\perp$ and $\mu_{0}H_\parallel$. The behavior of $\rho_{\mathrm{xx}}(T)$ for \#A and \#B under $\mu_{0}H_\perp$ = 0 T, 1 T, 3 T, 5 T and 7 T is provided in Figs.~\ref{fig:fig3}(a) and \ref{fig:fig3}(b), respectively. It is observed, that with the increase of $\mu_{0}H_\perp$, for both samples \#A and \#B  $T_\mathrm{loc}^\mathrm{onset}$ and $T_\mathrm{glo}^\mathrm{offset}$ are shifted to lower values compared to the ones in the ZFC case. The kink in the $\rho_{\mathrm{xx}}(T)$ detected in \#A vanishes upon the application of $\mu_{0}H_\perp$. For \#B, the anomalous peak in $\rho_{\mathrm{xx}}(T)$ diminishes in amplitude, $T_\mathrm{peak}$ shifts to lower $T$ and vanishes for $\mu_{0}H_\perp=3$~T. In addition, the resistivity minima at $T = T_\mathrm{d}$ and the kink at $T = T^{*}$ are also found to quench for $\mu_{0}H_\perp\ge3$~T. The resistivity $\rho_{\mathrm{xx}}(\mu_{0}H)$ of \#A and \#B is recorded at different $T$ for both $\mu_{0}H_\perp$ and $\mu_{0}H_\parallel$. In Figs.~\ref{fig:fig3}(c) and \ref{fig:fig3}(d), the $\rho_{\mathrm{xx}}(\mu_{0}H)$ of \#A and \#B measured for an applied $\mu_{0}H_\perp$ are presented. In agreement with the results of $\rho_{\mathrm{xx}}(T)$ measurements, the anomalous peak is also observed in the measured $\rho_{\mathrm{xx}}(\mu_{0}H)$ recorded for $\mu_{0}H_\perp$. The five characteristic temperatures observed in the ZFC $\rho_{\mathrm{xx}}(T)$ behavior of \#A and \#B along with $\Delta{T_\mathrm{c}}$ are summarized in Table 2.
	
	\begin{table}[]
		\caption{Characteristic temperatures of samples \#A and \#B recorded for ZFC $\rho_{\mathrm{xx}}(T)$.}
		\begin{tabular}{|l|l|l|l|l|l|l|}
			\hline
			Sample & $T_\mathrm{loc,ZFC}^\mathrm{onset}$ (K)    & $T_\mathrm{d,ZFC}$ (K)    & $T_\mathrm{peak,ZFC}$  (K)  & $T_\mathrm{ZFC}^{*}$   (K)  & $T_\mathrm{glo,ZFC}^\mathrm{offset}$ (K)  & $\Delta{\mathrm{T}}$ (K) \\ \hline
			A      & 15.72 & -     & -     & 15.27  & 15.05 & 0.67 \\ \hline
			B      & 15.1  & 14.71 & 14.60 & 14.305 & 13.5  & 1.6  \\ \hline
		\end{tabular}
		\label{tab:tab2}
	\end{table}

	
	The $\rho_{\mathrm{xx}}(T)$ and $\rho_{\mathrm{xx}}(\mu_{0}H)$ have been also measured for an applied $\mu_{0}H_\perp$. From the evolution of $\rho_{\mathrm{xx}}(T)$ and $\rho_{\mathrm{xx}}(\mu_{0}H)$ measured at different $T$ and as a function of $\mu_{0}H_\perp$ and $\mu_{0}H_\parallel$, the characteristic temperatures namely -  $T_\mathrm{loc,\perp/\parallel}^\mathrm{onset}$, $T_\mathrm{d,\perp/\parallel}$, $T_\mathrm{peak,\perp/\parallel}$, $T_{\perp/\parallel}^{*}$ and $T_\mathrm{glo,\perp/\parallel}^\mathrm{offset}$ - are estimated. The subscripts $\perp$ and $\parallel$ refer to $\mu_{0}H_\perp$ and $\mu_{0}H_\parallel$, respectively.
	\begin{figure*}[htbp]
		\centering
		\includegraphics[scale=1.20]{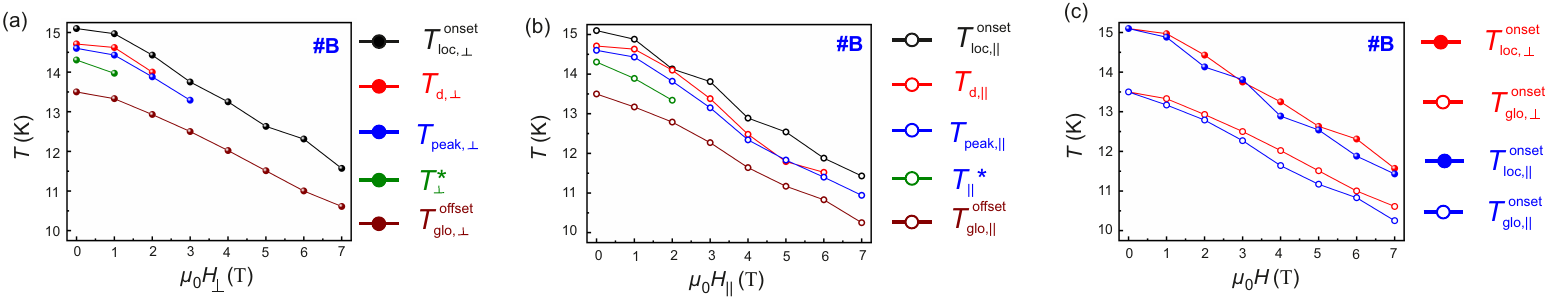}
		\caption{\label{fig:HT} The behavior of the characteristic temperatures  $T_\mathrm{loc}^\mathrm{onset}$, $T_\mathrm{d}$, $T_\mathrm{peak}$, $T = T^{*}$ and $T_\mathrm{glo}^\mathrm{offset}$ as a function of (a) $\mu_{0}H_\perp$. (b) $\mu_{0}H_\perp$. (c) Evolution of $T_\mathrm{loc,\perp/\parallel}^\mathrm{onset}$ recorded for \#B.}	
		\label{fig:fig4}
	\end{figure*}

	The behavior of the characteristic temperatures of \#B as a function of  $\mu_{0}H_\perp$ and $\mu_{0}H_\parallel$ is shown in 
	Figs.~\ref{fig:fig4}(a) and \ref{fig:fig4}(b), respectively. While $T_\mathrm{d,\parallel}$ and $T_\mathrm{peak,\parallel}$ are found to persist over the whole range of the applied $\mu_{0}H_\parallel$ ($i.e.$ $0\,\mathrm{T}\le{\mu_{0}H_\parallel}\le7\,\mathrm{T}$), $T_\mathrm{d,\perp}$ and $T_\mathrm{peak,\perp}$ are suppressed for $\mu_{0}H_\perp>{3\,\mathrm{T}}$. This anisotropy in the behavior of $T_\mathrm{d}$ and $T_\mathrm{peak}$ for magnetic fields applied parallel and perpendicular to the surface normal is not observed in similar disordered systems such as B-doped diamond \cite{Zhang:2013_PRL,Zhang:2016_PRApp} and $a$-InO \cite{Sacepe:2015_PRB} thin films. A comparison between $T_\mathrm{loc,\perp/\parallel}^\mathrm{onset}$ and $T_\mathrm{glo,\perp/\parallel}^\mathrm{offset}$ as a function of $\mu_{0}H$ for \#B is shown in Fig.~\ref{fig:fig4}(c). The similarity of these two characteristic temperatures for \#B indicates that the system is in a 3D regime and a 3D to 2D crossover of dimensionality, as the one reported for ultrathin TiN layers, can be ruled out \cite{Postolova:2017_SciRep}. The calculated full-width-and-half-maxima of $T_\mathrm{Peak}$ and $\Delta{T_\mathrm{c}}$ estimated for samples \#A and \#B recorded for $\mu_{0}H_{\parallel}$ and $\mu_{0}H_{\perp}$ are summarized in Table 3.
	
	Highly disordered superconducting thin films can show a reentrant resistive behavior, either due to a dimensionality crossover \cite{Postolova:2017_SciRep} or due to the application of a strong $\mu_{0}H$ \cite{Sacepe:2015_PRB}. The implantation of Fe into the NbN matrix enhances the disorder of the system compared to that of the as-grown NbN. The reentrant $T$ dependence of the normal state is referred to as a $N$-shaped $T$ dependence \cite{Postolova:2017_SciRep}, as the one recorded for the ZFC $\rho_{\mathrm{xx}}(T)$ of \#B. Such $N$-shaped $T$ dependence was reported in high $T_\mathrm{c}$ superconductors (HTS) \cite{Peng:2013_NatCom,Tagaki:1992_PRL,Ono:2000_PRL,Semba:2001_PRL,Komiya:2005_PRL,Oh:2006_PRL,Moshchalkov:2001_PRB} and ultrathin films of conventional superconductors \cite{Postolova:2017_SciRep,Sacepe:2015_PRB,Zhang:2013_PRL,Zhang:2016_PRApp}. Even though this phenomenon is ubiquitious, it still remains a subject of debate. In HTSs, this effect is described in terms of scaling functions \cite{Oh:2006_PRL,Moshchalkov:2001_PRB} or attributed to the emergence of the pseudogap phase \cite{Daou:2009_NatPhys}.


	\begin{table}[]
		\caption{The full width and half maxima of the $T_\mathrm{Peak}$ and $\Delta{T_\mathrm{c}}$ estimated for samples \#A and \#B recorded for $\mu_{0}H_{\parallel}$ and $\mu_{0}H_{\perp}$.}
		\begin{tabular}{|l|ll|ll|ll|ll|}
			\hline
			$\mu_{0}H_{\parallel/\perp}$ (T) & \multicolumn{2}{l|}{$\Delta{\mathrm{T}_\mathrm{Peak}^\mathrm{A}} (\mathrm{K})$}    & \multicolumn{2}{l|}{$\Delta{\mathrm{T}_\mathrm{Peak}^\mathrm{B}} (\mathrm{K})$}         & \multicolumn{2}{l|}{$\Delta{\mathrm{T}_\mathrm{c}^\mathrm{A}} (\mathrm{K})$}      & \multicolumn{2}{l|}{$\Delta{\mathrm{T}_\mathrm{c}^\mathrm{B}} (\mathrm{K})$}      \\ \hline
			& \multicolumn{1}{l|}{$\mu_{0}H_{\perp}$} & {$\mu_{0}H_{\parallel}$} & \multicolumn{1}{l|}{$\mu_{0}H_{\perp}$}   & {$\mu_{0}H_{\parallel}$}    & \multicolumn{1}{l|}{$\mu_{0}H_{\perp}$} & {$\mu_{0}H_{\parallel}$}  & \multicolumn{1}{l|}{{$\mu_{0}H_{\parallel}$}} & {$\mu_{0}H_{\parallel}$}  \\ \hline
			0     & \multicolumn{1}{l|}{-}    & -   & \multicolumn{1}{l|}{0.0653} & 0.0653 & \multicolumn{1}{l|}{0.67} & 0.67 & \multicolumn{1}{l|}{1.6}  & 1.6  \\ \hline
			1     & \multicolumn{1}{l|}{-}    & -   & \multicolumn{1}{l|}{0.1376} & 0.1337 & \multicolumn{1}{l|}{0.72} & 0.94 & \multicolumn{1}{l|}{1.64} & 1.71 \\ \hline
			2     & \multicolumn{1}{l|}{-}    & -   & \multicolumn{1}{l|}{0.0858} & 0.1514 & \multicolumn{1}{l|}{-}    & -    & \multicolumn{1}{l|}{1.5}  & 1.34 \\ \hline
			3     & \multicolumn{1}{l|}{-}    & -   & \multicolumn{1}{l|}{-}      & 0.1392 & \multicolumn{1}{l|}{0.74} & 0.94 & \multicolumn{1}{l|}{1.25} & 1.54 \\ \hline
			4     & \multicolumn{1}{l|}{-}    & -   & \multicolumn{1}{l|}{-}      & 0.0841 & \multicolumn{1}{l|}{-}    & -    & \multicolumn{1}{l|}{1.23} & 1.25 \\ \hline
			5     & \multicolumn{1}{l|}{-}    & -   & \multicolumn{1}{l|}{-}      & -      & \multicolumn{1}{l|}{0.8}  & 0.96 & \multicolumn{1}{l|}{1.12} & 1.37 \\ \hline
			6     & \multicolumn{1}{l|}{-}    & -   & \multicolumn{1}{l|}{-}      & -      & \multicolumn{1}{l|}{-}    & -    & \multicolumn{1}{l|}{1.31} & 1.05 \\ \hline
			7     & \multicolumn{1}{l|}{-}    & -   & \multicolumn{1}{l|}{-}      & -      & \multicolumn{1}{l|}{0.78} & 0.91 & \multicolumn{1}{l|}{0.96} & 1.18 \\ \hline
		\end{tabular}
		\label{tab:tab3}
	\end{table}

	
	\begin{figure*}[htbp]
		\centering
		\includegraphics[scale=1.75]{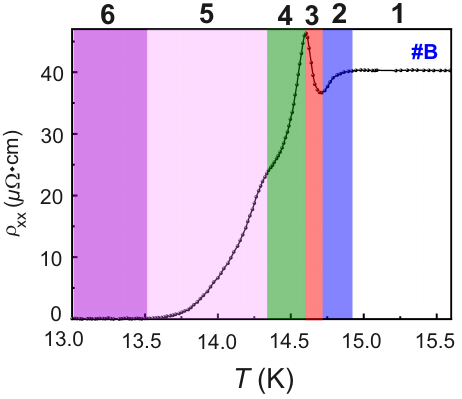}
		\caption{\label{fig:ZFCM} ZFC $\rho_{\mathrm{xx}}$ as a function of $T$. The six electonic phases and the phase boundaries of \#B. are evidenced}	
		\label{fig:fig5}
	\end{figure*}

	In the case of ultrathin $<15$ nm layers of conventional superconductors, quantum contributions due to a 3D to 2D dimensionality crossover have been identified as the mechanism behind the observed $N$-shaped $T$ dependence of $\rho_{\mathrm{xx}}$. In 3D films of disordered granular superconductors such as B-doped diamond \cite{Zhang:2013_PRL,Zhang:2016_PRApp}, $a$-InO \cite{Sacepe:2015_PRB} or AlGe \cite{Zaken:1994_JPCM} this behavior is interpreted within the framework of a N-BI-S transition originating from granularity induced disorder.
	
	In the ZFC $\rho_{\mathrm{xx}}(T)$ of \#B a series of electronic phases can be identified. These electronic phases are marked by EP-1, EP-2, EP-3. EP-4, EP-5 and EP-6 in Fig.~\ref{fig:fig5}. The phase boundaries of the adjacent electronic phases of \#B correspond to the characteristic temperatures $T_\mathrm{loc}^\mathrm{onset}$, $T_\mathrm{d}$, $T_\mathrm{peak}$, $T^{*}$ and $T_\mathrm{glo}^\mathrm{offset}$, as described earlier. 
	
\begin{figure*}[htbp]
		\centering
		\includegraphics[scale=1.50]{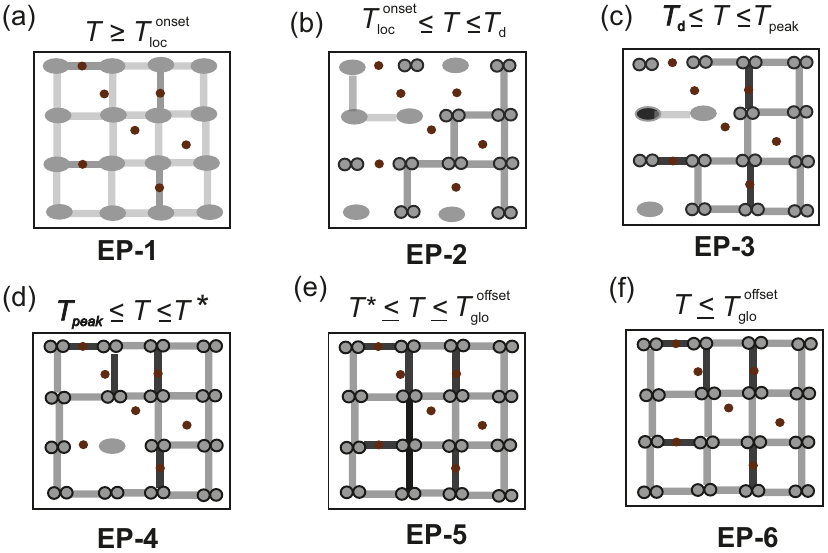}
		\caption{\label{fig:SCHM} Schematic of the empirical model to describe the N-BI-S transition in \#B. The connector lines connecting the ellipses and dumbbells represent fermionic conduction channels and bosonic conduction channels, respectively. The normal state fermions and the Cooper pairs are shown by ellipses and dumbbells, respectively, while the Fe impurities are depicted by solid circles. The electronic phase transitions taking place as the sample \#B is cooled from $T>T_\mathrm{loc}^\mathrm{onset}$ down to $T\le{T_\mathrm{loc}^\mathrm{onset}}$ are: (a) Phase EP-1 for $T>T_\mathrm{loc}^\mathrm{onset}$. (b) Phase EP-2 for $T_\mathrm{d}<T<T_\mathrm{loc}^\mathrm{onset}$. (c) Phase EP-3 for $T^\mathrm{peak}<T<T_\mathrm{d}$. (d) Phase EP-4 for $T_\mathrm{Peak}\le{T}\le{T^{*}}$. (e) Phase EP-5 for $T^{*}\le{T}\le{T_\mathrm{glo}^\mathrm{offset}}$ and (f) Phase EP-6 for $T\le{T_\mathrm{glo}^\mathrm{offset}}$.}	
		\label{fig:fig6}
	\end{figure*}

	The evolution of the six electronic phases, $i.e.$ from EP-1 to EP-6, across the phase boundaries is explained using an empirical model of competing impurity scattering and a conduction network of fermionic conduction channel and bosonic conduction channels  across a N-BI-S phase transition. A schematic of this empirical model is presented in Fig.~\ref{fig:fig6}. The Fe implantation dose of $\left(1\times10^{14}\right)$ at/cm$^3$ results in a highly dilute Fe doped system, restricting the Fe-Fe interaction to the paramagnetic limit. Since after implantation \#B is not thermally treated, the Fe ions are incorporated into the NbN lattice randomly. Moreover, the incorporation efficiency of implanted Fe ions is different for NbN grains with different crystallographic orientations. The disorder of the Fe implanted NbN system stems from an extrinsic granularity and an intrinsic one. The disorder due to the polycrystalline texture and grain boundaries of the Fe:NbN lattice is the extrinsic granularity. On the other hand, the electronic disorder due to the randomly implanted Fe ions results in modulations of the chemical potential and thus in intrinsic electronic granularity. The intrinsic granularity plays a significant role in defining the observed electronic properties of \#B.

	The mechanisms leading to the specific electronic phases are:
	
	(i) Phase EP-1: The normal electronic phase in the ZFC $\rho_{\mathrm{xx}}(T)$ of Fe:NbN sets in for $T>T_\mathrm{loc}^\mathrm{onset}$, as shown in Fig.~\ref{fig:fig5}. This electronic phase is represented in Fig.~\ref{fig:fig6}(a) and is referred to as EP-1. For $T>T_\mathrm{loc}^\mathrm{onset}$, the electrical transport takes place $via$ the percolation of thermally activated fermionic conduction channels with normal electrons being the medium of transport. These fermionic conduction channels are marked in the schematic diagram as solid lines and the fermions are visualized by solid ellipses. The Fe impurities in the matrix are represented by solid circles. The resistivity behavior of this phase is the same as the one of the N phase of Fe:NbN.
	
	(ii) Phase EP-2: As $T$ decreases, a superconducting gap opens at $T=T_\mathrm{loc}^\mathrm{onset}$. A spatial inhomogeneity associated with the superconducting gap due to the intrinsic granularity leads to the formation of bosonic clusters. With a gradual decrease of $T$ below $T=T_\mathrm{loc}^\mathrm{onset}$, the density of the bosonic clusters increases and becomes locally phase locked, resulting in the initial decrease of $\rho_{\mathrm{xx}}$ with decreasing $T$ through the percolation of the phase locked bosonic clusters merging into bosonic islands. These islands are responsible for the bosonic conduction channels for electronic transport in the temperature range $T_\mathrm{d}<T<T_\mathrm{loc}^\mathrm{onset}$. The bosonic conduction channels and the bosonic islands are represented in Fig.~\ref{fig:fig6}(b) by connectors and dumb bells, respectively. In this regime, a competition of the bosonic conduction channels with the fermionic conduction channels leads to a decrease in $\rho_{\mathrm{xx}}$.
	
	(iii) Phase EP-3: As the $T$ is further reduced, $i.e.$ for $T<T_\mathrm{d}$, in the presence of the intrinsic granularity due to the paramagnetic Fe impurities, the bosonic conduction channels dominate over the fermionic conduction channels $via$ the removal of single fermions or normal electrons from the fermionic conduction channels to fuel the bosonic conduction channels. This process takes place by forming groups of phase-locked bosonic clusters and producing isolated BIs. The consequent anomalous increase of $\rho_{\mathrm{xx}}$ for $T^\mathrm{peak}<T<T_\mathrm{d}$ is likely due to two mechanisms, $i.e.$ 
	
	- removal of the single fermions from the fermionic conduction channels to estblish the percolation path for the bosonic conduction channels;
	
	- scattering of the bosonic clusters from the isolated and randomly distributed Fe ions in the matrix.  
	
	The anomalous increase of $\rho_{\mathrm{xx}}$ continues up to $T=T_\mathrm{peak}$ The electronic phase EP-3 is referred to as a bosonic insulator phase and is represented in Fig.~\ref{fig:fig6}(c). 
	
	(iv) Phase EP-4: As $T$ is further lowered, the density of the bosonic conduction channels increases $via$ the percolation of more bosonic islands, leading to a decrease in $\rho_{\mathrm{xx}}$, as more and more phase locked coherent bosonic clusters and islands percolate in the bulk of the system. The dissipationless transport due to the coherent bosonic islands dominates over the scattering from the paramagnetic Fe impurities, leading to the gradual decrease of $\rho_{\mathrm{xx}}$ until $T = T^{*}$. The EP-4 phase is shown in Fig.~\ref{fig:fig6}(d).
	
	(v) Phase EP-5: At $T = T^{*}$, a kink in $\rho_{\mathrm{xx}}(T)$ is observed, which is attributed to the scattering of the percolating bosonic islands due the extrinsic granularity. This is also detected in the as-grown NbN \#A, as shown in Fig.~\ref{fig:fig6}(e).
	
	(vi) Phase EP-6: At $T\le{T_\mathrm{loc}^\mathrm{onset}}$, the global phase coherence sets in between all the phase-locked bosonic islands and the electrical transport is dominated solely by the bosonic conduction channels, leading to a superconducting state of the Fe implanted NbN layer, as sketched in Fig.~\ref{fig:fig6}(f). 
	
	The empirical model also describes the behavior of $\rho_{\mathrm{xx}}(T)$ for applied $\mu_{0}H_\perp$ and $\mu_{0}H_\parallel$. The presence of $\mu_{0}H_\perp$ breaks the Cooper pairing and shifts $T_\mathrm{loc}^\mathrm{onset}$ to a lower $T$. This also suppresses the onset of the bosonic insulator phase. In addition, the formation of the Abrikosov vortices for $\mu_{0}H_\perp$ also plays a role in the suppression of the BI phase and of the intrinsic granularity on the electronic transport of \#B.
	
	For $\mu_{0}H_\parallel$ however, the fields required to break the Cooper pairs are generally orders of magnitude stronger than the ones for $\mu_{0}H_\perp$ and do not influence the formation of the BI state. As a result both $T^\mathrm{peak,\parallel}$and $T_\mathrm{d,\parallel}$ persist up to the highest applied field  $\mu_{0}H_\parallel=7\,\mathrm{T}$.

	
	\section{Conclusions}
	
	In conclusion, 100 nm thick crystalline films of NbN are deposited by reactive magnetron sputtering on single crystalline GaN templates grown on epi ready $c$-plane sapphire substrates. Paramagnetic Fe doping of the NbN layers is achieved by ion implantation using Fe$_{3}$O$_{4}$ as the source for Fe ions. An implantation energy of 35 keV and a dose of $\left(1\times10^{14}\right)$ at/cm$^3$ result in a highly dilute Fe doped system, restricting the Fe-Fe interaction to the paramagnetic limit. Low-$T$/high-$\mu_{0}H$ magnetotransport measurements confirm, that the Fe doping does not suppress the superconductivity of the sputtered NbN thin films, but decreases the superconducting transition temperature. A reentrant resistive BI phase is observed in the Fe doped NbN samples, which is explained by an empirical model of a competition between the percolation of bosonic conduction channels at the expense of fermionic conduction channels and by the scattering of the BI as a result of intrinsic granularity due to the random Fe dopants. The observation of a robust superconductivity in the dilute magnetic conventional superconductor Fe:NbN mediated $via$ percolation of bosonic insulator states is foreseen to violate the symmetry of electron-like and hole-like excitations due to the formation of subgap bound Andreev states in the vicinity of magnetic impurities, leading to giant thermoelectric effects \cite{Kalenkov:2012_PRL}. A system like the one reported in this work is expected to find applications in zero biased thermoelectric bolometers with reduced power dissipation in large scale multi-pixel arrays and in hybrid quantum interference devices (HyQUID) \cite{Shelley:2021_SciAdv}. Moreover, these systems are the workbench for understanding quantum emergent phenomena, including gapless superconductivity, triplet Cooper pairings, YSR states and odd frequency superconductivity.

	\section*{Acknowledgements}
	
	The work was funded by the Austrian Science Fund (FWF) through Projects No. P26830 and No. P31423.

	\bibliographystyle{apsrev4-2}
	
	%

	


\begin{thebibliography}{64}%
		\makeatletter
		\providecommand \@ifxundefined [1]{%
			\@ifx{#1\undefined}
		}%
		\providecommand \@ifnum [1]{%
			\ifnum #1\expandafter \@firstoftwo
			\else \expandafter \@secondoftwo
			\fi
		}%
		\providecommand \@ifx [1]{%
			\ifx #1\expandafter \@firstoftwo
			\else \expandafter \@secondoftwo
			\fi
		}%
		\providecommand \natexlab [1]{#1}%
		\providecommand \enquote  [1]{``#1''}%
		\providecommand \bibnamefont  [1]{#1}%
		\providecommand \bibfnamefont [1]{#1}%
		\providecommand \citenamefont [1]{#1}%
		\providecommand \href@noop [0]{\@secondoftwo}%
		\providecommand \href [0]{\begingroup \@sanitize@url \@href}%
		\providecommand \@href[1]{\@@startlink{#1}\@@href}%
		\providecommand \@@href[1]{\endgroup#1\@@endlink}%
		\providecommand \@sanitize@url [0]{\catcode `\\12\catcode `\$12\catcode
			`\&12\catcode `\#12\catcode `\^12\catcode `\_12\catcode `\%12\relax}%
		\providecommand \@@startlink[1]{}%
		\providecommand \@@endlink[0]{}%
		\providecommand \url  [0]{\begingroup\@sanitize@url \@url }%
		\providecommand \@url [1]{\endgroup\@href {#1}{\urlprefix }}%
		\providecommand \urlprefix  [0]{URL }%
		\providecommand \Eprint [0]{\href }%
		\providecommand \doibase [0]{https://doi.org/}%
		\providecommand \selectlanguage [0]{\@gobble}%
		\providecommand \bibinfo  [0]{\@secondoftwo}%
		\providecommand \bibfield  [0]{\@secondoftwo}%
		\providecommand \translation [1]{[#1]}%
		\providecommand \BibitemOpen [0]{}%
		\providecommand \bibitemStop [0]{}%
		\providecommand \bibitemNoStop [0]{.\EOS\space}%
		\providecommand \EOS [0]{\spacefactor3000\relax}%
		\providecommand \BibitemShut  [1]{\csname bibitem#1\endcsname}%
		\let\auto@bib@innerbib\@empty
		\bibitem [{\citenamefont {Balatsky}\ \emph {et~al.}(2006)\citenamefont
			{Balatsky}, \citenamefont {Vekhter},\ and\ \citenamefont
			{Zhu}}]{Balatsky:2006_RMP}%
		\BibitemOpen
		\bibfield  {author} {\bibinfo {author} {\bibfnamefont {A.~V.}\ \bibnamefont
				{Balatsky}}, \bibinfo {author} {\bibfnamefont {I.}~\bibnamefont {Vekhter}},\
			and\ \bibinfo {author} {\bibfnamefont {J.-X.}\ \bibnamefont {Zhu}},\ }\href
		{https://doi.org/10.1103/RevModPhys.78.373} {\bibfield  {journal} {\bibinfo
				{journal} {Rev. Mod. Phys.}\ }\textbf {\bibinfo {volume} {78}},\ \bibinfo
			{pages} {373} (\bibinfo {year} {2006})}\BibitemShut {NoStop}%
		\bibitem [{\citenamefont {M{\"o}ckli}\ \emph {et~al.}(2020)\citenamefont
			{M{\"o}ckli}, \citenamefont {Haim},\ and\ \citenamefont
			{Khodas}}]{Mockli:2020_JAP}%
		\BibitemOpen
		\bibfield  {author} {\bibinfo {author} {\bibfnamefont {D.}~\bibnamefont
				{M{\"o}ckli}}, \bibinfo {author} {\bibfnamefont {M.}~\bibnamefont {Haim}},\
			and\ \bibinfo {author} {\bibfnamefont {M.}~\bibnamefont {Khodas}},\ }\href
		{https://doi.org/10.1063/5.0010773} {\bibfield  {journal} {\bibinfo
				{journal} {J. Appl. Phys}\ }\textbf {\bibinfo {volume} {128}},\ \bibinfo
			{pages} {053903} (\bibinfo {year} {2020})}\BibitemShut {NoStop}%
		\bibitem [{\citenamefont {Gor'kov}(2008)}]{Gorkov:2008_Book}%
		\BibitemOpen
		\bibfield  {author} {\bibinfo {author} {\bibfnamefont {L.~P.}\ \bibnamefont
				{Gor'kov}},\ }\bibinfo {title} {Theory of superconducting alloys},\ in\ \href
		{https://doi.org/10.1007/978-3-540-73253-2_5} {\emph {\bibinfo {booktitle}
				{Superconductivity: Conventional and Unconventional Superconductors}}}\
		(\bibinfo  {publisher} {Springer Berlin Heidelberg},\ \bibinfo {address}
		{Berlin, Heidelberg},\ \bibinfo {year} {2008})\ pp.\ \bibinfo {pages}
		{201--224}\BibitemShut {NoStop}%
		\bibitem [{\citenamefont {Anderson}(1959)}]{Anderson:1959_JPCS}%
		\BibitemOpen
		\bibfield  {author} {\bibinfo {author} {\bibfnamefont {P.~W.}\ \bibnamefont
				{Anderson}},\ }\href
		{https://www.sciencedirect.com/science/article/pii/0022369759900368}
		{\bibfield  {journal} {\bibinfo  {journal} {J Phys. Chem. Solids}\ }\textbf
			{\bibinfo {volume} {11}},\ \bibinfo {pages} {26} (\bibinfo {year}
			{1959})}\BibitemShut {NoStop}%
		\bibitem [{\citenamefont {Abrikosov}\ and\ \citenamefont
			{Gor'kov}(1960)}]{Abrikosov:1961_JETP}%
		\BibitemOpen
		\bibfield  {author} {\bibinfo {author} {\bibfnamefont {A.~A.}\ \bibnamefont
				{Abrikosov}}\ and\ \bibinfo {author} {\bibfnamefont {L.~P.}\ \bibnamefont
				{Gor'kov}},\ }\bibfield  {journal} {\bibinfo  {journal} {Zhur. Eksptl'. i
				Teoret. Fiz.}\ }\textbf {\bibinfo {volume} {39}},\ \href
		{https://www.osti.gov/biblio/4097498} {} (\bibinfo {year} {1960})\BibitemShut
		{NoStop}%
		\bibitem [{\citenamefont {Abrikosov}(1969)}]{Abrikosov:1969_JETP}%
		\BibitemOpen
		\bibfield  {author} {\bibinfo {author} {\bibfnamefont {A.~A.}\ \bibnamefont
				{Abrikosov}},\ }\href {https://doi.org/10.1070/pu1969v012n02abeh003930}
		{\bibfield  {journal} {\bibinfo  {journal} {Sov. Phys. Usp.}\ }\textbf
			{\bibinfo {volume} {12}},\ \bibinfo {pages} {168} (\bibinfo {year}
			{1969})}\BibitemShut {NoStop}%
		\bibitem [{\citenamefont {Zittartz}\ and\ \citenamefont
			{M{\"u}ller-Hartmann}(1970)}]{Zittartz:1970_ZPhys_1}%
		\BibitemOpen
		\bibfield  {author} {\bibinfo {author} {\bibfnamefont {J.}~\bibnamefont
				{Zittartz}}\ and\ \bibinfo {author} {\bibfnamefont {E.}~\bibnamefont
				{M{\"u}ller-Hartmann}},\ }\href {https://doi.org/10.1007/BF01394943}
		{\bibfield  {journal} {\bibinfo  {journal} {Zeitschrift f{\"u}r Physik A
					Hadrons and nuclei}\ }\textbf {\bibinfo {volume} {232}},\ \bibinfo {pages}
			{11} (\bibinfo {year} {1970})}\BibitemShut {NoStop}%
		\bibitem [{\citenamefont {M{\"u}ller-Hartmann}\ and\ \citenamefont
			{Zittartz}(1970)}]{Mueller-Hartmann:1970_ZPhys}%
		\BibitemOpen
		\bibfield  {author} {\bibinfo {author} {\bibfnamefont {E.}~\bibnamefont
				{M{\"u}ller-Hartmann}}\ and\ \bibinfo {author} {\bibfnamefont
				{J.}~\bibnamefont {Zittartz}},\ }\href {https://doi.org/10.1007/BF01392497}
		{\bibfield  {journal} {\bibinfo  {journal} {Zeitschrift f{\"u}r Physik A
					Hadrons and nuclei}\ }\textbf {\bibinfo {volume} {234}},\ \bibinfo {pages}
			{58} (\bibinfo {year} {1970})}\BibitemShut {NoStop}%
		\bibitem [{\citenamefont {Zittartz}(1970)}]{Zittartz:1970_ZPhys_2}%
		\BibitemOpen
		\bibfield  {author} {\bibinfo {author} {\bibfnamefont {J.}~\bibnamefont
				{Zittartz}},\ }\href {https://doi.org/10.1007/BF01407639} {\bibfield
			{journal} {\bibinfo  {journal} {Zeitschrift f{\"u}r Physik A Hadrons and
					nuclei}\ }\textbf {\bibinfo {volume} {237}},\ \bibinfo {pages} {419}
			(\bibinfo {year} {1970})}\BibitemShut {NoStop}%
		\bibitem [{\citenamefont {Woolf}\ and\ \citenamefont
			{Reif}(1965)}]{Woolf:1965_PhysRev}%
		\BibitemOpen
		\bibfield  {author} {\bibinfo {author} {\bibfnamefont {M.~A.}\ \bibnamefont
				{Woolf}}\ and\ \bibinfo {author} {\bibfnamefont {F.}~\bibnamefont {Reif}},\
		}\href {https://doi.org/10.1103/PhysRev.137.A557} {\bibfield  {journal}
			{\bibinfo  {journal} {Phys. Rev.}\ }\textbf {\bibinfo {volume} {137}},\
			\bibinfo {pages} {A557} (\bibinfo {year} {1965})}\BibitemShut {NoStop}%
		\bibitem [{\citenamefont {Skalski}\ \emph {et~al.}(1964)\citenamefont
			{Skalski}, \citenamefont {Betbeder-Matibet},\ and\ \citenamefont
			{Weiss}}]{Skalski:1964_PhysRev}%
		\BibitemOpen
		\bibfield  {author} {\bibinfo {author} {\bibfnamefont {S.}~\bibnamefont
				{Skalski}}, \bibinfo {author} {\bibfnamefont {O.}~\bibnamefont
				{Betbeder-Matibet}},\ and\ \bibinfo {author} {\bibfnamefont {P.~R.}\
				\bibnamefont {Weiss}},\ }\href {https://doi.org/10.1103/PhysRev.136.A1500}
		{\bibfield  {journal} {\bibinfo  {journal} {Phys. Rev.}\ }\textbf {\bibinfo
				{volume} {136}},\ \bibinfo {pages} {A1500} (\bibinfo {year}
			{1964})}\BibitemShut {NoStop}%
		\bibitem [{\citenamefont {Fulde}\ and\ \citenamefont
			{Maki}(1966)}]{Fulde:1966_PhysRev}%
		\BibitemOpen
		\bibfield  {author} {\bibinfo {author} {\bibfnamefont {P.}~\bibnamefont
				{Fulde}}\ and\ \bibinfo {author} {\bibfnamefont {K.}~\bibnamefont {Maki}},\
		}\href {https://doi.org/10.1103/PhysRev.141.275} {\bibfield  {journal}
			{\bibinfo  {journal} {Phys. Rev.}\ }\textbf {\bibinfo {volume} {141}},\
			\bibinfo {pages} {275} (\bibinfo {year} {1966})}\BibitemShut {NoStop}%
		\bibitem [{\citenamefont {Fulde}(1973)}]{Fulde:1973_Adv.Phys.}%
		\BibitemOpen
		\bibfield  {author} {\bibinfo {author} {\bibfnamefont {P.}~\bibnamefont
				{Fulde}},\ }\href {https://doi.org/10.1080/00018737300101369} {\bibfield
			{journal} {\bibinfo  {journal} {Advances in Physics}\ }\textbf {\bibinfo
				{volume} {22}},\ \bibinfo {pages} {667} (\bibinfo {year} {1973})}\BibitemShut
		{NoStop}%
		\bibitem [{\citenamefont {Maki}(1967)}]{Maki:1967_PhysRev}%
		\BibitemOpen
		\bibfield  {author} {\bibinfo {author} {\bibfnamefont {K.}~\bibnamefont
				{Maki}},\ }\href {https://doi.org/10.1103/PhysRev.153.428} {\bibfield
			{journal} {\bibinfo  {journal} {Phys. Rev.}\ }\textbf {\bibinfo {volume}
				{153}},\ \bibinfo {pages} {428} (\bibinfo {year} {1967})}\BibitemShut
		{NoStop}%
		\bibitem [{\citenamefont {Cooper}(1956)}]{Cooper:1956_PhysRev}%
		\BibitemOpen
		\bibfield  {author} {\bibinfo {author} {\bibfnamefont {L.~N.}\ \bibnamefont
				{Cooper}},\ }\href {https://doi.org/10.1103/PhysRev.104.1189} {\bibfield
			{journal} {\bibinfo  {journal} {Phys. Rev.}\ }\textbf {\bibinfo {volume}
				{104}},\ \bibinfo {pages} {1189} (\bibinfo {year} {1956})}\BibitemShut
		{NoStop}%
		\bibitem [{\citenamefont {Bardeen}\ \emph
			{et~al.}(1957{\natexlab{a}})\citenamefont {Bardeen}, \citenamefont {Cooper},\
			and\ \citenamefont {Schrieffer}}]{BCS:1957_PhysRev_1}%
		\BibitemOpen
		\bibfield  {author} {\bibinfo {author} {\bibfnamefont {J.}~\bibnamefont
				{Bardeen}}, \bibinfo {author} {\bibfnamefont {L.~N.}\ \bibnamefont
				{Cooper}},\ and\ \bibinfo {author} {\bibfnamefont {J.~R.}\ \bibnamefont
				{Schrieffer}},\ }\href {https://doi.org/10.1103/PhysRev.106.162} {\bibfield
			{journal} {\bibinfo  {journal} {Phys. Rev.}\ }\textbf {\bibinfo {volume}
				{106}},\ \bibinfo {pages} {162} (\bibinfo {year}
			{1957}{\natexlab{a}})}\BibitemShut {NoStop}%
		\bibitem [{\citenamefont {Bardeen}\ \emph
			{et~al.}(1957{\natexlab{b}})\citenamefont {Bardeen}, \citenamefont {Cooper},\
			and\ \citenamefont {Schrieffer}}]{BCS:1957_PhysRev_2}%
		\BibitemOpen
		\bibfield  {author} {\bibinfo {author} {\bibfnamefont {J.}~\bibnamefont
				{Bardeen}}, \bibinfo {author} {\bibfnamefont {L.~N.}\ \bibnamefont
				{Cooper}},\ and\ \bibinfo {author} {\bibfnamefont {J.~R.}\ \bibnamefont
				{Schrieffer}},\ }\href {https://doi.org/10.1103/PhysRev.108.1175} {\bibfield
			{journal} {\bibinfo  {journal} {Phys. Rev.}\ }\textbf {\bibinfo {volume}
				{108}},\ \bibinfo {pages} {1175} (\bibinfo {year}
			{1957}{\natexlab{b}})}\BibitemShut {NoStop}%
		\bibitem [{\citenamefont {Pint}\ and\ \citenamefont
			{Schachinger}(1989)}]{Pint:1989_PhysC}%
		\BibitemOpen
		\bibfield  {author} {\bibinfo {author} {\bibfnamefont {W.}~\bibnamefont
				{Pint}}\ and\ \bibinfo {author} {\bibfnamefont {E.}~\bibnamefont
				{Schachinger}},\ }\href
		{https://www.sciencedirect.com/science/article/pii/0921453489901007}
		{\bibfield  {journal} {\bibinfo  {journal} {Physica C: Superconductivity}\
			}\textbf {\bibinfo {volume} {159}},\ \bibinfo {pages} {33} (\bibinfo {year}
			{1989})}\BibitemShut {NoStop}%
		\bibitem [{\citenamefont {Heinrich}\ \emph {et~al.}(2018)\citenamefont
			{Heinrich}, \citenamefont {Pascual},\ and\ \citenamefont
			{Franke}}]{Heinrich:2018_PSS}%
		\BibitemOpen
		\bibfield  {author} {\bibinfo {author} {\bibfnamefont {B.~W.}\ \bibnamefont
				{Heinrich}}, \bibinfo {author} {\bibfnamefont {J.~I.}\ \bibnamefont
				{Pascual}},\ and\ \bibinfo {author} {\bibfnamefont {K.~J.}\ \bibnamefont
				{Franke}},\ }\href
		{https://www.sciencedirect.com/science/article/pii/S0079681618300017}
		{\bibfield  {journal} {\bibinfo  {journal} {Prog. Surf. Sci}\ }\textbf
			{\bibinfo {volume} {93}},\ \bibinfo {pages} {1} (\bibinfo {year}
			{2018})}\BibitemShut {NoStop}%
		\bibitem [{\citenamefont {Linder}\ and\ \citenamefont
			{Robinson}(2015)}]{Linder:2015_NatPhys}%
		\BibitemOpen
		\bibfield  {author} {\bibinfo {author} {\bibfnamefont {J.}~\bibnamefont
				{Linder}}\ and\ \bibinfo {author} {\bibfnamefont {J.~W.~A.}\ \bibnamefont
				{Robinson}},\ }\href {https://doi.org/10.1038/nphys3242} {\bibfield
			{journal} {\bibinfo  {journal} {Nat. Phys.}\ }\textbf {\bibinfo {volume}
				{11}},\ \bibinfo {pages} {307} (\bibinfo {year} {2015})}\BibitemShut
		{NoStop}%
		\bibitem [{\citenamefont {M\"uller}\ \emph {et~al.}(2021)\citenamefont
			{M\"uller}, \citenamefont {Liensberger}, \citenamefont {Flacke},
			\citenamefont {Huebl}, \citenamefont {Kamra}, \citenamefont {Belzig},
			\citenamefont {Gross}, \citenamefont {Weiler},\ and\ \citenamefont
			{Althammer}}]{Mueller:2021_PRL}%
		\BibitemOpen
		\bibfield  {author} {\bibinfo {author} {\bibfnamefont {M.}~\bibnamefont
				{M\"uller}}, \bibinfo {author} {\bibfnamefont {L.}~\bibnamefont
				{Liensberger}}, \bibinfo {author} {\bibfnamefont {L.}~\bibnamefont {Flacke}},
			\bibinfo {author} {\bibfnamefont {H.}~\bibnamefont {Huebl}}, \bibinfo
			{author} {\bibfnamefont {A.}~\bibnamefont {Kamra}}, \bibinfo {author}
			{\bibfnamefont {W.}~\bibnamefont {Belzig}}, \bibinfo {author} {\bibfnamefont
				{R.}~\bibnamefont {Gross}}, \bibinfo {author} {\bibfnamefont
				{M.}~\bibnamefont {Weiler}},\ and\ \bibinfo {author} {\bibfnamefont
				{M.}~\bibnamefont {Althammer}},\ }\href
		{https://doi.org/10.1103/PhysRevLett.126.087201} {\bibfield  {journal}
			{\bibinfo  {journal} {Phys. Rev. Lett.}\ }\textbf {\bibinfo {volume} {126}},\
			\bibinfo {pages} {087201} (\bibinfo {year} {2021})}\BibitemShut {NoStop}%
		\bibitem [{\citenamefont {Fulde}\ and\ \citenamefont
			{Ferrell}(1964)}]{Fulde:1964_PhysRev}%
		\BibitemOpen
		\bibfield  {author} {\bibinfo {author} {\bibfnamefont {P.}~\bibnamefont
				{Fulde}}\ and\ \bibinfo {author} {\bibfnamefont {R.~A.}\ \bibnamefont
				{Ferrell}},\ }\href {https://doi.org/10.1103/PhysRev.135.A550} {\bibfield
			{journal} {\bibinfo  {journal} {Phys. Rev.}\ }\textbf {\bibinfo {volume}
				{135}},\ \bibinfo {pages} {A550} (\bibinfo {year} {1964})}\BibitemShut
		{NoStop}%
		\bibitem [{\citenamefont {Larkin}\ and\ \citenamefont
			{Ovchinnikov}(1964)}]{Larkin:1964_JETP}%
		\BibitemOpen
		\bibfield  {author} {\bibinfo {author} {\bibfnamefont {A.~I.}\ \bibnamefont
				{Larkin}}\ and\ \bibinfo {author} {\bibfnamefont {Y.~N.}\ \bibnamefont
				{Ovchinnikov}},\ }\bibfield  {journal} {\bibinfo  {journal} {Zh. Eksperim. i
				Teor. Fiz.}\ }\textbf {\bibinfo {volume} {47}},\ \href
		{https://www.osti.gov/biblio/4653415} {} (\bibinfo {year} {1964})\BibitemShut
		{NoStop}%
		\bibitem [{\citenamefont {Lenk}\ \emph {et~al.}(2016)\citenamefont {Lenk},
			\citenamefont {Hemmida}, \citenamefont {Morari}, \citenamefont {Zdravkov},
			\citenamefont {Ullrich}, \citenamefont {M\"uller}, \citenamefont {Sidorenko},
			\citenamefont {Horn}, \citenamefont {Tagirov}, \citenamefont {Loidl},
			\citenamefont {von Nidda},\ and\ \citenamefont {Tidecks}}]{Lenk:2016_PRB}%
		\BibitemOpen
		\bibfield  {author} {\bibinfo {author} {\bibfnamefont {D.}~\bibnamefont
				{Lenk}}, \bibinfo {author} {\bibfnamefont {M.}~\bibnamefont {Hemmida}},
			\bibinfo {author} {\bibfnamefont {R.}~\bibnamefont {Morari}}, \bibinfo
			{author} {\bibfnamefont {V.~I.}\ \bibnamefont {Zdravkov}}, \bibinfo {author}
			{\bibfnamefont {A.}~\bibnamefont {Ullrich}}, \bibinfo {author} {\bibfnamefont
				{C.}~\bibnamefont {M\"uller}}, \bibinfo {author} {\bibfnamefont {A.~S.}\
				\bibnamefont {Sidorenko}}, \bibinfo {author} {\bibfnamefont {S.}~\bibnamefont
				{Horn}}, \bibinfo {author} {\bibfnamefont {L.~R.}\ \bibnamefont {Tagirov}},
			\bibinfo {author} {\bibfnamefont {A.}~\bibnamefont {Loidl}}, \bibinfo
			{author} {\bibfnamefont {H.-A.~K.}\ \bibnamefont {von Nidda}},\ and\ \bibinfo
			{author} {\bibfnamefont {R.}~\bibnamefont {Tidecks}},\ }\href
		{https://doi.org/10.1103/PhysRevB.93.184501} {\bibfield  {journal} {\bibinfo
				{journal} {Phys. Rev. B}\ }\textbf {\bibinfo {volume} {93}},\ \bibinfo
			{pages} {184501} (\bibinfo {year} {2016})}\BibitemShut {NoStop}%
		\bibitem [{\citenamefont {Buzdin}(2005)}]{Buzdin:2005_RMP}%
		\BibitemOpen
		\bibfield  {author} {\bibinfo {author} {\bibfnamefont {A.~I.}\ \bibnamefont
				{Buzdin}},\ }\href {https://doi.org/10.1103/RevModPhys.77.935} {\bibfield
			{journal} {\bibinfo  {journal} {Rev. Mod. Phys.}\ }\textbf {\bibinfo {volume}
				{77}},\ \bibinfo {pages} {935} (\bibinfo {year} {2005})}\BibitemShut
		{NoStop}%
		\bibitem [{\citenamefont {Chockalingam}\ \emph {et~al.}(2008)\citenamefont
			{Chockalingam}, \citenamefont {Chand}, \citenamefont {Jesudasan},
			\citenamefont {Tripathi},\ and\ \citenamefont
			{Raychaudhuri}}]{Chockalingam:2008_PRB}%
		\BibitemOpen
		\bibfield  {author} {\bibinfo {author} {\bibfnamefont {S.~P.}\ \bibnamefont
				{Chockalingam}}, \bibinfo {author} {\bibfnamefont {M.}~\bibnamefont {Chand}},
			\bibinfo {author} {\bibfnamefont {J.}~\bibnamefont {Jesudasan}}, \bibinfo
			{author} {\bibfnamefont {V.}~\bibnamefont {Tripathi}},\ and\ \bibinfo
			{author} {\bibfnamefont {P.}~\bibnamefont {Raychaudhuri}},\ }\href
		{https://doi.org/10.1103/PhysRevB.77.214503} {\bibfield  {journal} {\bibinfo
				{journal} {Phys. Rev. B}\ }\textbf {\bibinfo {volume} {77}},\ \bibinfo
			{pages} {214503} (\bibinfo {year} {2008})}\BibitemShut {NoStop}%
		\bibitem [{\citenamefont {Koushik}\ \emph {et~al.}(2013)\citenamefont
			{Koushik}, \citenamefont {Kumar}, \citenamefont {Amin}, \citenamefont
			{Mondal}, \citenamefont {Jesudasan}, \citenamefont {Bid}, \citenamefont
			{Raychaudhuri},\ and\ \citenamefont {Ghosh}}]{Koushik:2013_PRL}%
		\BibitemOpen
		\bibfield  {author} {\bibinfo {author} {\bibfnamefont {R.}~\bibnamefont
				{Koushik}}, \bibinfo {author} {\bibfnamefont {S.}~\bibnamefont {Kumar}},
			\bibinfo {author} {\bibfnamefont {K.~R.}\ \bibnamefont {Amin}}, \bibinfo
			{author} {\bibfnamefont {M.}~\bibnamefont {Mondal}}, \bibinfo {author}
			{\bibfnamefont {J.}~\bibnamefont {Jesudasan}}, \bibinfo {author}
			{\bibfnamefont {A.}~\bibnamefont {Bid}}, \bibinfo {author} {\bibfnamefont
				{P.}~\bibnamefont {Raychaudhuri}},\ and\ \bibinfo {author} {\bibfnamefont
				{A.}~\bibnamefont {Ghosh}},\ }\href
		{https://doi.org/10.1103/PhysRevLett.111.197001} {\bibfield  {journal}
			{\bibinfo  {journal} {Phys. Rev. Lett.}\ }\textbf {\bibinfo {volume} {111}},\
			\bibinfo {pages} {197001} (\bibinfo {year} {2013})}\BibitemShut {NoStop}%
		\bibitem [{\citenamefont {Ganguly}\ \emph {et~al.}(2015)\citenamefont
			{Ganguly}, \citenamefont {Chaudhuri}, \citenamefont {Raychaudhuri},\ and\
			\citenamefont {Benfatto}}]{Ganguly:2015_PRB}%
		\BibitemOpen
		\bibfield  {author} {\bibinfo {author} {\bibfnamefont {R.}~\bibnamefont
				{Ganguly}}, \bibinfo {author} {\bibfnamefont {D.}~\bibnamefont {Chaudhuri}},
			\bibinfo {author} {\bibfnamefont {P.}~\bibnamefont {Raychaudhuri}},\ and\
			\bibinfo {author} {\bibfnamefont {L.}~\bibnamefont {Benfatto}},\ }\href
		{https://doi.org/10.1103/PhysRevB.91.054514} {\bibfield  {journal} {\bibinfo
				{journal} {Phys. Rev. B}\ }\textbf {\bibinfo {volume} {91}},\ \bibinfo
			{pages} {054514} (\bibinfo {year} {2015})}\BibitemShut {NoStop}%
		\bibitem [{\citenamefont {Chand}\ \emph {et~al.}(2009)\citenamefont {Chand},
			\citenamefont {Mishra}, \citenamefont {Xiong}, \citenamefont {Kamlapure},
			\citenamefont {Chockalingam}, \citenamefont {Jesudasan}, \citenamefont
			{Bagwe}, \citenamefont {Mondal}, \citenamefont {Adams}, \citenamefont
			{Tripathi},\ and\ \citenamefont {Raychaudhuri}}]{Chand:2009_PRB}%
		\BibitemOpen
		\bibfield  {author} {\bibinfo {author} {\bibfnamefont {M.}~\bibnamefont
				{Chand}}, \bibinfo {author} {\bibfnamefont {A.}~\bibnamefont {Mishra}},
			\bibinfo {author} {\bibfnamefont {Y.~M.}\ \bibnamefont {Xiong}}, \bibinfo
			{author} {\bibfnamefont {A.}~\bibnamefont {Kamlapure}}, \bibinfo {author}
			{\bibfnamefont {S.~P.}\ \bibnamefont {Chockalingam}}, \bibinfo {author}
			{\bibfnamefont {J.}~\bibnamefont {Jesudasan}}, \bibinfo {author}
			{\bibfnamefont {V.}~\bibnamefont {Bagwe}}, \bibinfo {author} {\bibfnamefont
				{M.}~\bibnamefont {Mondal}}, \bibinfo {author} {\bibfnamefont {P.~W.}\
				\bibnamefont {Adams}}, \bibinfo {author} {\bibfnamefont {V.}~\bibnamefont
				{Tripathi}},\ and\ \bibinfo {author} {\bibfnamefont {P.}~\bibnamefont
				{Raychaudhuri}},\ }\href {https://doi.org/10.1103/PhysRevB.80.134514}
		{\bibfield  {journal} {\bibinfo  {journal} {Phys. Rev. B}\ }\textbf {\bibinfo
				{volume} {80}},\ \bibinfo {pages} {134514} (\bibinfo {year}
			{2009})}\BibitemShut {NoStop}%
		\bibitem [{\citenamefont {Chand}\ \emph {et~al.}(2012)\citenamefont {Chand},
			\citenamefont {Saraswat}, \citenamefont {Kamlapure}, \citenamefont {Mondal},
			\citenamefont {Kumar}, \citenamefont {Jesudasan}, \citenamefont {Bagwe},
			\citenamefont {Benfatto}, \citenamefont {Tripathi},\ and\ \citenamefont
			{Raychaudhuri}}]{Chand:2012_PRB}%
		\BibitemOpen
		\bibfield  {author} {\bibinfo {author} {\bibfnamefont {M.}~\bibnamefont
				{Chand}}, \bibinfo {author} {\bibfnamefont {G.}~\bibnamefont {Saraswat}},
			\bibinfo {author} {\bibfnamefont {A.}~\bibnamefont {Kamlapure}}, \bibinfo
			{author} {\bibfnamefont {M.}~\bibnamefont {Mondal}}, \bibinfo {author}
			{\bibfnamefont {S.}~\bibnamefont {Kumar}}, \bibinfo {author} {\bibfnamefont
				{J.}~\bibnamefont {Jesudasan}}, \bibinfo {author} {\bibfnamefont
				{V.}~\bibnamefont {Bagwe}}, \bibinfo {author} {\bibfnamefont
				{L.}~\bibnamefont {Benfatto}}, \bibinfo {author} {\bibfnamefont
				{V.}~\bibnamefont {Tripathi}},\ and\ \bibinfo {author} {\bibfnamefont
				{P.}~\bibnamefont {Raychaudhuri}},\ }\href
		{https://doi.org/10.1103/PhysRevB.85.014508} {\bibfield  {journal} {\bibinfo
				{journal} {Phys. Rev. B}\ }\textbf {\bibinfo {volume} {85}},\ \bibinfo
			{pages} {014508} (\bibinfo {year} {2012})}\BibitemShut {NoStop}%
		\bibitem [{\citenamefont {Mondal}\ \emph
			{et~al.}(2011{\natexlab{a}})\citenamefont {Mondal}, \citenamefont {Kumar},
			\citenamefont {Chand}, \citenamefont {Kamlapure}, \citenamefont {Saraswat},
			\citenamefont {Seibold}, \citenamefont {Benfatto},\ and\ \citenamefont
			{Raychaudhuri}}]{Mondal:2011_PRL}%
		\BibitemOpen
		\bibfield  {author} {\bibinfo {author} {\bibfnamefont {M.}~\bibnamefont
				{Mondal}}, \bibinfo {author} {\bibfnamefont {S.}~\bibnamefont {Kumar}},
			\bibinfo {author} {\bibfnamefont {M.}~\bibnamefont {Chand}}, \bibinfo
			{author} {\bibfnamefont {A.}~\bibnamefont {Kamlapure}}, \bibinfo {author}
			{\bibfnamefont {G.}~\bibnamefont {Saraswat}}, \bibinfo {author}
			{\bibfnamefont {G.}~\bibnamefont {Seibold}}, \bibinfo {author} {\bibfnamefont
				{L.}~\bibnamefont {Benfatto}},\ and\ \bibinfo {author} {\bibfnamefont
				{P.}~\bibnamefont {Raychaudhuri}},\ }\href
		{https://doi.org/10.1103/PhysRevLett.107.217003} {\bibfield  {journal}
			{\bibinfo  {journal} {Phys. Rev. Lett.}\ }\textbf {\bibinfo {volume} {107}},\
			\bibinfo {pages} {217003} (\bibinfo {year} {2011}{\natexlab{a}})}\BibitemShut
		{NoStop}%
		\bibitem [{\citenamefont {Chockalingam}\ \emph {et~al.}(2009)\citenamefont
			{Chockalingam}, \citenamefont {Chand}, \citenamefont {Kamlapure},
			\citenamefont {Jesudasan}, \citenamefont {Mishra}, \citenamefont {Tripathi},\
			and\ \citenamefont {Raychaudhuri}}]{Chockalingam:2009_PRB}%
		\BibitemOpen
		\bibfield  {author} {\bibinfo {author} {\bibfnamefont {S.~P.}\ \bibnamefont
				{Chockalingam}}, \bibinfo {author} {\bibfnamefont {M.}~\bibnamefont {Chand}},
			\bibinfo {author} {\bibfnamefont {A.}~\bibnamefont {Kamlapure}}, \bibinfo
			{author} {\bibfnamefont {J.}~\bibnamefont {Jesudasan}}, \bibinfo {author}
			{\bibfnamefont {A.}~\bibnamefont {Mishra}}, \bibinfo {author} {\bibfnamefont
				{V.}~\bibnamefont {Tripathi}},\ and\ \bibinfo {author} {\bibfnamefont
				{P.}~\bibnamefont {Raychaudhuri}},\ }\href
		{https://doi.org/10.1103/PhysRevB.79.094509} {\bibfield  {journal} {\bibinfo
				{journal} {Phys. Rev. B}\ }\textbf {\bibinfo {volume} {79}},\ \bibinfo
			{pages} {094509} (\bibinfo {year} {2009})}\BibitemShut {NoStop}%
		\bibitem [{\citenamefont {Destraz}\ \emph {et~al.}(2017)\citenamefont
			{Destraz}, \citenamefont {Ilin}, \citenamefont {Siegel}, \citenamefont
			{Schilling},\ and\ \citenamefont {Chang}}]{Destraz:2017_PRB}%
		\BibitemOpen
		\bibfield  {author} {\bibinfo {author} {\bibfnamefont {D.}~\bibnamefont
				{Destraz}}, \bibinfo {author} {\bibfnamefont {K.}~\bibnamefont {Ilin}},
			\bibinfo {author} {\bibfnamefont {M.}~\bibnamefont {Siegel}}, \bibinfo
			{author} {\bibfnamefont {A.}~\bibnamefont {Schilling}},\ and\ \bibinfo
			{author} {\bibfnamefont {J.}~\bibnamefont {Chang}},\ }\href
		{https://doi.org/10.1103/PhysRevB.95.224501} {\bibfield  {journal} {\bibinfo
				{journal} {Phys. Rev. B}\ }\textbf {\bibinfo {volume} {95}},\ \bibinfo
			{pages} {224501} (\bibinfo {year} {2017})}\BibitemShut {NoStop}%
		\bibitem [{\citenamefont {Nikzad}\ \emph {et~al.}(2016)\citenamefont {Nikzad},
			\citenamefont {Hoenk}, \citenamefont {Jewell}, \citenamefont {Hennessy},
			\citenamefont {Carver}, \citenamefont {Jones}, \citenamefont {Goodsall},
			\citenamefont {Hamden}, \citenamefont {Suvarna}, \citenamefont {Bulmer},
			\citenamefont {Shahedipour-Sandvik}, \citenamefont {Charbon}, \citenamefont
			{Padmanabhan}, \citenamefont {Hancock},\ and\ \citenamefont
			{Bell}}]{Nikzad:2016_Sensors}%
		\BibitemOpen
		\bibfield  {author} {\bibinfo {author} {\bibfnamefont {S.}~\bibnamefont
				{Nikzad}}, \bibinfo {author} {\bibfnamefont {M.}~\bibnamefont {Hoenk}},
			\bibinfo {author} {\bibfnamefont {A.~D.}\ \bibnamefont {Jewell}}, \bibinfo
			{author} {\bibfnamefont {J.~J.}\ \bibnamefont {Hennessy}}, \bibinfo {author}
			{\bibfnamefont {A.~G.}\ \bibnamefont {Carver}}, \bibinfo {author}
			{\bibfnamefont {T.~J.}\ \bibnamefont {Jones}}, \bibinfo {author}
			{\bibfnamefont {T.~M.}\ \bibnamefont {Goodsall}}, \bibinfo {author}
			{\bibfnamefont {E.~T.}\ \bibnamefont {Hamden}}, \bibinfo {author}
			{\bibfnamefont {P.}~\bibnamefont {Suvarna}}, \bibinfo {author} {\bibfnamefont
				{J.}~\bibnamefont {Bulmer}}, \bibinfo {author} {\bibfnamefont
				{F.}~\bibnamefont {Shahedipour-Sandvik}}, \bibinfo {author} {\bibfnamefont
				{E.}~\bibnamefont {Charbon}}, \bibinfo {author} {\bibfnamefont
				{P.}~\bibnamefont {Padmanabhan}}, \bibinfo {author} {\bibfnamefont
				{B.}~\bibnamefont {Hancock}},\ and\ \bibinfo {author} {\bibfnamefont {L.~D.}\
				\bibnamefont {Bell}},\ }\bibfield  {journal} {\bibinfo  {journal} {Sensors}\
		}\textbf {\bibinfo {volume} {16}},\ \href {https://doi.org/10.3390/s16060927}
		{10.3390/s16060927} (\bibinfo {year} {2016})\BibitemShut {NoStop}%
		\bibitem [{\citenamefont {Polakovic}\ \emph {et~al.}(2020)\citenamefont
			{Polakovic}, \citenamefont {Armstrong}, \citenamefont {Karapetrov},
			\citenamefont {Meziani},\ and\ \citenamefont
			{Novosad}}]{Polakovic:2020_Nanomater}%
		\BibitemOpen
		\bibfield  {author} {\bibinfo {author} {\bibfnamefont {T.}~\bibnamefont
				{Polakovic}}, \bibinfo {author} {\bibfnamefont {W.}~\bibnamefont
				{Armstrong}}, \bibinfo {author} {\bibfnamefont {G.}~\bibnamefont
				{Karapetrov}}, \bibinfo {author} {\bibfnamefont {Z.-E.}\ \bibnamefont
				{Meziani}},\ and\ \bibinfo {author} {\bibfnamefont {V.}~\bibnamefont
				{Novosad}},\ }\bibfield  {journal} {\bibinfo  {journal} {Nanomaterials}\
		}\textbf {\bibinfo {volume} {10}},\ \href
		{https://doi.org/10.3390/nano10061198} {10.3390/nano10061198} (\bibinfo
		{year} {2020})\BibitemShut {NoStop}%
		\bibitem [{\citenamefont {Blais}\ \emph {et~al.}(2021)\citenamefont {Blais},
			\citenamefont {Grimsmo}, \citenamefont {Girvin},\ and\ \citenamefont
			{Wallraff}}]{Blais:2021_RMP}%
		\BibitemOpen
		\bibfield  {author} {\bibinfo {author} {\bibfnamefont {A.}~\bibnamefont
				{Blais}}, \bibinfo {author} {\bibfnamefont {A.~L.}\ \bibnamefont {Grimsmo}},
			\bibinfo {author} {\bibfnamefont {S.~M.}\ \bibnamefont {Girvin}},\ and\
			\bibinfo {author} {\bibfnamefont {A.}~\bibnamefont {Wallraff}},\ }\href
		{https://doi.org/10.1103/RevModPhys.93.025005} {\bibfield  {journal}
			{\bibinfo  {journal} {Rev. Mod. Phys.}\ }\textbf {\bibinfo {volume} {93}},\
			\bibinfo {pages} {025005} (\bibinfo {year} {2021})}\BibitemShut {NoStop}%
		\bibitem [{\citenamefont {Yong}\ \emph {et~al.}(2013)\citenamefont {Yong},
			\citenamefont {Lemberger}, \citenamefont {Benfatto}, \citenamefont {Ilin},\
			and\ \citenamefont {Siegel}}]{Yong:2013_PRB}%
		\BibitemOpen
		\bibfield  {author} {\bibinfo {author} {\bibfnamefont {J.}~\bibnamefont
				{Yong}}, \bibinfo {author} {\bibfnamefont {T.~R.}\ \bibnamefont {Lemberger}},
			\bibinfo {author} {\bibfnamefont {L.}~\bibnamefont {Benfatto}}, \bibinfo
			{author} {\bibfnamefont {K.}~\bibnamefont {Ilin}},\ and\ \bibinfo {author}
			{\bibfnamefont {M.}~\bibnamefont {Siegel}},\ }\href
		{https://doi.org/10.1103/PhysRevB.87.184505} {\bibfield  {journal} {\bibinfo
				{journal} {Phys. Rev. B}\ }\textbf {\bibinfo {volume} {87}},\ \bibinfo
			{pages} {184505} (\bibinfo {year} {2013})}\BibitemShut {NoStop}%
		\bibitem [{\citenamefont {Mondal}\ \emph
			{et~al.}(2011{\natexlab{b}})\citenamefont {Mondal}, \citenamefont {Chand},
			\citenamefont {Kamlapure}, \citenamefont {Jesudasan}, \citenamefont {Bagwe},
			\citenamefont {Kumar}, \citenamefont {Saraswat}, \citenamefont {Tripathi},\
			and\ \citenamefont {Raychaudhuri}}]{Mondal:2011_JSNM}%
		\BibitemOpen
		\bibfield  {author} {\bibinfo {author} {\bibfnamefont {M.}~\bibnamefont
				{Mondal}}, \bibinfo {author} {\bibfnamefont {M.}~\bibnamefont {Chand}},
			\bibinfo {author} {\bibfnamefont {A.}~\bibnamefont {Kamlapure}}, \bibinfo
			{author} {\bibfnamefont {J.}~\bibnamefont {Jesudasan}}, \bibinfo {author}
			{\bibfnamefont {V.~C.}\ \bibnamefont {Bagwe}}, \bibinfo {author}
			{\bibfnamefont {S.}~\bibnamefont {Kumar}}, \bibinfo {author} {\bibfnamefont
				{G.}~\bibnamefont {Saraswat}}, \bibinfo {author} {\bibfnamefont
				{V.}~\bibnamefont {Tripathi}},\ and\ \bibinfo {author} {\bibfnamefont
				{P.}~\bibnamefont {Raychaudhuri}},\ }\href
		{https://doi.org/10.1007/s10948-010-1038-8} {\bibfield  {journal} {\bibinfo
				{journal} {J Supercond. Nov. Magn.}\ }\textbf {\bibinfo {volume} {24}},\
			\bibinfo {pages} {341} (\bibinfo {year} {2011}{\natexlab{b}})}\BibitemShut
		{NoStop}%
		\bibitem [{\citenamefont {Sherman}\ \emph {et~al.}(2015)\citenamefont
			{Sherman}, \citenamefont {Pracht}, \citenamefont {Gorshunov}, \citenamefont
			{Poran}, \citenamefont {Jesudasan}, \citenamefont {Chand}, \citenamefont
			{Raychaudhuri}, \citenamefont {Swanson}, \citenamefont {Trivedi},
			\citenamefont {Auerbach}, \citenamefont {Scheffler}, \citenamefont
			{Frydman},\ and\ \citenamefont {Dressel}}]{Sherman:2015_NatPhys}%
		\BibitemOpen
		\bibfield  {author} {\bibinfo {author} {\bibfnamefont {D.}~\bibnamefont
				{Sherman}}, \bibinfo {author} {\bibfnamefont {U.~S.}\ \bibnamefont {Pracht}},
			\bibinfo {author} {\bibfnamefont {B.}~\bibnamefont {Gorshunov}}, \bibinfo
			{author} {\bibfnamefont {S.}~\bibnamefont {Poran}}, \bibinfo {author}
			{\bibfnamefont {J.}~\bibnamefont {Jesudasan}}, \bibinfo {author}
			{\bibfnamefont {M.}~\bibnamefont {Chand}}, \bibinfo {author} {\bibfnamefont
				{P.}~\bibnamefont {Raychaudhuri}}, \bibinfo {author} {\bibfnamefont
				{M.}~\bibnamefont {Swanson}}, \bibinfo {author} {\bibfnamefont
				{N.}~\bibnamefont {Trivedi}}, \bibinfo {author} {\bibfnamefont
				{A.}~\bibnamefont {Auerbach}}, \bibinfo {author} {\bibfnamefont
				{M.}~\bibnamefont {Scheffler}}, \bibinfo {author} {\bibfnamefont
				{A.}~\bibnamefont {Frydman}},\ and\ \bibinfo {author} {\bibfnamefont
				{M.}~\bibnamefont {Dressel}},\ }\href {https://doi.org/10.1038/nphys3227}
		{\bibfield  {journal} {\bibinfo  {journal} {Nat. Phys.}\ }\textbf {\bibinfo
				{volume} {11}},\ \bibinfo {pages} {188} (\bibinfo {year} {2015})}\BibitemShut
		{NoStop}%
		\bibitem [{\citenamefont {Tsuji}\ and\ \citenamefont
			{Nomura}(2020)}]{Tsuji:2020_PRRes}%
		\BibitemOpen
		\bibfield  {author} {\bibinfo {author} {\bibfnamefont {N.}~\bibnamefont
				{Tsuji}}\ and\ \bibinfo {author} {\bibfnamefont {Y.}~\bibnamefont {Nomura}},\
		}\href {https://doi.org/10.1103/PhysRevResearch.2.043029} {\bibfield
			{journal} {\bibinfo  {journal} {Phys. Rev. Research}\ }\textbf {\bibinfo
				{volume} {2}},\ \bibinfo {pages} {043029} (\bibinfo {year}
			{2020})}\BibitemShut {NoStop}%
		\bibitem [{\citenamefont {Hochberg}\ \emph {et~al.}(2016)\citenamefont
			{Hochberg}, \citenamefont {Zhao},\ and\ \citenamefont
			{Zurek}}]{Hochberg:2016_PRL}%
		\BibitemOpen
		\bibfield  {author} {\bibinfo {author} {\bibfnamefont {Y.}~\bibnamefont
				{Hochberg}}, \bibinfo {author} {\bibfnamefont {Y.}~\bibnamefont {Zhao}},\
			and\ \bibinfo {author} {\bibfnamefont {K.~M.}\ \bibnamefont {Zurek}},\ }\href
		{https://doi.org/10.1103/PhysRevLett.116.011301} {\bibfield  {journal}
			{\bibinfo  {journal} {Phys. Rev. Lett.}\ }\textbf {\bibinfo {volume} {116}},\
			\bibinfo {pages} {011301} (\bibinfo {year} {2016})}\BibitemShut {NoStop}%
		\bibitem [{\citenamefont {Krause}\ \emph {et~al.}(2014)\citenamefont {Krause},
			\citenamefont {Meledin}, \citenamefont {Desmaris}, \citenamefont
			{Pavolotsky}, \citenamefont {Belitsky}, \citenamefont {Rudzi{\'{n}}ski},\
			and\ \citenamefont {Pippel}}]{Krause:2014_SST}%
		\BibitemOpen
		\bibfield  {author} {\bibinfo {author} {\bibfnamefont {S.}~\bibnamefont
				{Krause}}, \bibinfo {author} {\bibfnamefont {D.}~\bibnamefont {Meledin}},
			\bibinfo {author} {\bibfnamefont {V.}~\bibnamefont {Desmaris}}, \bibinfo
			{author} {\bibfnamefont {A.}~\bibnamefont {Pavolotsky}}, \bibinfo {author}
			{\bibfnamefont {V.}~\bibnamefont {Belitsky}}, \bibinfo {author}
			{\bibfnamefont {M.}~\bibnamefont {Rudzi{\'{n}}ski}},\ and\ \bibinfo {author}
			{\bibfnamefont {E.}~\bibnamefont {Pippel}},\ }\href
		{https://doi.org/10.1088/0953-2048/27/6/065009} {\bibfield  {journal}
			{\bibinfo  {journal} {Supercond. Sci. Technol}\ }\textbf {\bibinfo {volume}
				{27}},\ \bibinfo {pages} {065009} (\bibinfo {year} {2014})}\BibitemShut
		{NoStop}%
		\bibitem [{\citenamefont {Sam-Giao}\ \emph {et~al.}(2014)\citenamefont
			{Sam-Giao}, \citenamefont {Pouget}, \citenamefont {Bougerol}, \citenamefont
			{Monroy}, \citenamefont {Grimm}, \citenamefont {Jebari}, \citenamefont
			{Hofheinz}, \citenamefont {G{\'e}rard},\ and\ \citenamefont
			{Zwiller}}]{Sam-Giao:2014_AIPAdv}%
		\BibitemOpen
		\bibfield  {author} {\bibinfo {author} {\bibfnamefont {D.}~\bibnamefont
				{Sam-Giao}}, \bibinfo {author} {\bibfnamefont {S.}~\bibnamefont {Pouget}},
			\bibinfo {author} {\bibfnamefont {C.}~\bibnamefont {Bougerol}}, \bibinfo
			{author} {\bibfnamefont {E.}~\bibnamefont {Monroy}}, \bibinfo {author}
			{\bibfnamefont {A.}~\bibnamefont {Grimm}}, \bibinfo {author} {\bibfnamefont
				{S.}~\bibnamefont {Jebari}}, \bibinfo {author} {\bibfnamefont
				{M.}~\bibnamefont {Hofheinz}}, \bibinfo {author} {\bibfnamefont {J.-M.}\
				\bibnamefont {G{\'e}rard}},\ and\ \bibinfo {author} {\bibfnamefont
				{V.}~\bibnamefont {Zwiller}},\ }\href {https://doi.org/10.1063/1.4898327}
		{\bibfield  {journal} {\bibinfo  {journal} {AIP Adv.}\ }\textbf {\bibinfo
				{volume} {4}},\ \bibinfo {pages} {107123} (\bibinfo {year}
			{2014})}\BibitemShut {NoStop}%
		\bibitem [{\citenamefont {Kobayashi}\ \emph
			{et~al.}(2020{\natexlab{a}})\citenamefont {Kobayashi}, \citenamefont {Ueno},\
			and\ \citenamefont {Fujioka}}]{Kobayashi:2020_APE}%
		\BibitemOpen
		\bibfield  {author} {\bibinfo {author} {\bibfnamefont {A.}~\bibnamefont
				{Kobayashi}}, \bibinfo {author} {\bibfnamefont {K.}~\bibnamefont {Ueno}},\
			and\ \bibinfo {author} {\bibfnamefont {H.}~\bibnamefont {Fujioka}},\ }\href
		{https://doi.org/10.35848/1882-0786/ab916e} {\bibfield  {journal} {\bibinfo
				{journal} {Appl. Phys. Express}\ }\textbf {\bibinfo {volume} {13}},\ \bibinfo
			{pages} {061006} (\bibinfo {year} {2020}{\natexlab{a}})}\BibitemShut
		{NoStop}%
		\bibitem [{\citenamefont {Kobayashi}\ \emph
			{et~al.}(2020{\natexlab{b}})\citenamefont {Kobayashi}, \citenamefont {Ueno},\
			and\ \citenamefont {Fujioka}}]{Kobayashi:2020_APL}%
		\BibitemOpen
		\bibfield  {author} {\bibinfo {author} {\bibfnamefont {A.}~\bibnamefont
				{Kobayashi}}, \bibinfo {author} {\bibfnamefont {K.}~\bibnamefont {Ueno}},\
			and\ \bibinfo {author} {\bibfnamefont {H.}~\bibnamefont {Fujioka}},\ }\href
		{https://doi.org/10.1063/5.0031604} {\bibfield  {journal} {\bibinfo
				{journal} {App. Phys. Lett.}\ }\textbf {\bibinfo {volume} {117}},\ \bibinfo
			{pages} {231601} (\bibinfo {year} {2020}{\natexlab{b}})}\BibitemShut
		{NoStop}%
		\bibitem [{\citenamefont {Yan}\ \emph {et~al.}(2018)\citenamefont {Yan},
			\citenamefont {Khalsa}, \citenamefont {Vishwanath}, \citenamefont {Han},
			\citenamefont {Wright}, \citenamefont {Rouvimov}, \citenamefont {Katzer},
			\citenamefont {Nepal}, \citenamefont {Downey}, \citenamefont {Muller},
			\citenamefont {Xing}, \citenamefont {Meyer},\ and\ \citenamefont
			{Jena}}]{Yan:2018_Nature}%
		\BibitemOpen
		\bibfield  {author} {\bibinfo {author} {\bibfnamefont {R.}~\bibnamefont
				{Yan}}, \bibinfo {author} {\bibfnamefont {G.}~\bibnamefont {Khalsa}},
			\bibinfo {author} {\bibfnamefont {S.}~\bibnamefont {Vishwanath}}, \bibinfo
			{author} {\bibfnamefont {Y.}~\bibnamefont {Han}}, \bibinfo {author}
			{\bibfnamefont {J.}~\bibnamefont {Wright}}, \bibinfo {author} {\bibfnamefont
				{S.}~\bibnamefont {Rouvimov}}, \bibinfo {author} {\bibfnamefont {D.~S.}\
				\bibnamefont {Katzer}}, \bibinfo {author} {\bibfnamefont {N.}~\bibnamefont
				{Nepal}}, \bibinfo {author} {\bibfnamefont {B.~P.}\ \bibnamefont {Downey}},
			\bibinfo {author} {\bibfnamefont {D.~A.}\ \bibnamefont {Muller}}, \bibinfo
			{author} {\bibfnamefont {H.~G.}\ \bibnamefont {Xing}}, \bibinfo {author}
			{\bibfnamefont {D.~J.}\ \bibnamefont {Meyer}},\ and\ \bibinfo {author}
			{\bibfnamefont {D.}~\bibnamefont {Jena}},\ }\href
		{https://doi.org/10.1038/nature25768} {\bibfield  {journal} {\bibinfo
				{journal} {Nature}\ }\textbf {\bibinfo {volume} {555}},\ \bibinfo {pages}
			{183} (\bibinfo {year} {2018})}\BibitemShut {NoStop}%
		\bibitem [{\citenamefont {Hwang}\ and\ \citenamefont
			{Kim}(2017)}]{Hwang:2017_PhysC}%
		\BibitemOpen
		\bibfield  {author} {\bibinfo {author} {\bibfnamefont {T.-J.}\ \bibnamefont
				{Hwang}}\ and\ \bibinfo {author} {\bibfnamefont {D.-H.}\ \bibnamefont
				{Kim}},\ }\href
		{https://www.sciencedirect.com/science/article/pii/S0921453417301260}
		{\bibfield  {journal} {\bibinfo  {journal} {Physica C Supercond.}\ }\textbf
			{\bibinfo {volume} {540}},\ \bibinfo {pages} {16} (\bibinfo {year}
			{2017})}\BibitemShut {NoStop}%
		\bibitem [{\citenamefont {Jha}\ \emph {et~al.}(2013)\citenamefont {Jha},
			\citenamefont {Jyoti},\ and\ \citenamefont {Awana}}]{Jha:2013_JSNM}%
		\BibitemOpen
		\bibfield  {author} {\bibinfo {author} {\bibfnamefont {R.}~\bibnamefont
				{Jha}}, \bibinfo {author} {\bibfnamefont {J.}~\bibnamefont {Jyoti}},\ and\
			\bibinfo {author} {\bibfnamefont {V.~P.~S.}\ \bibnamefont {Awana}},\ }\href
		{https://doi.org/10.1007/s10948-013-2132-5} {\bibfield  {journal} {\bibinfo
				{journal} {J. Supercond. Nov. Magn.}\ }\textbf {\bibinfo {volume} {26}},\
			\bibinfo {pages} {3069} (\bibinfo {year} {2013})}\BibitemShut {NoStop}%
		\bibitem [{\citenamefont {Vorhauer}(2021)}]{Vorhauer:2021_Thesis}%
		\BibitemOpen
		\bibfield  {author} {\bibinfo {author} {\bibfnamefont {J.~V.}\ \bibnamefont
				{Vorhauer}},\ }\emph {\bibinfo {title} {Influence of Fe implantation on the
				superconductivity of NbN}},\ \href {https://permalink.obvsg.at/AC16384491}
		{Ph.D. thesis},\ \bibinfo  {school} {Johannes Kepler University, Linz}
		(\bibinfo {year} {2021})\BibitemShut {NoStop}%
		\bibitem [{\citenamefont {Zhang}\ \emph {et~al.}(2013)\citenamefont {Zhang},
			\citenamefont {Zeleznik}, \citenamefont {Vanacken}, \citenamefont {May},\
			and\ \citenamefont {Moshchalkov}}]{Zhang:2013_PRL}%
		\BibitemOpen
		\bibfield  {author} {\bibinfo {author} {\bibfnamefont {G.}~\bibnamefont
				{Zhang}}, \bibinfo {author} {\bibfnamefont {M.}~\bibnamefont {Zeleznik}},
			\bibinfo {author} {\bibfnamefont {J.}~\bibnamefont {Vanacken}}, \bibinfo
			{author} {\bibfnamefont {P.~W.}\ \bibnamefont {May}},\ and\ \bibinfo {author}
			{\bibfnamefont {V.~V.}\ \bibnamefont {Moshchalkov}},\ }\href
		{https://doi.org/10.1103/PhysRevLett.110.077001} {\bibfield  {journal}
			{\bibinfo  {journal} {Phys. Rev. Lett.}\ }\textbf {\bibinfo {volume} {110}},\
			\bibinfo {pages} {077001} (\bibinfo {year} {2013})}\BibitemShut {NoStop}%
		\bibitem [{\citenamefont {Zhang}\ \emph {et~al.}(2016)\citenamefont {Zhang},
			\citenamefont {Samuely}, \citenamefont {Ka\ifmmode \check{c}\else
				\v{c}\fi{}mar\ifmmode~\check{c}\else \v{c}\fi{}\'{\i}k}, \citenamefont
			{Ekimov}, \citenamefont {Li}, \citenamefont {Vanacken}, \citenamefont
			{Szab\'o}, \citenamefont {Huang}, \citenamefont {Pereira}, \citenamefont
			{Cerbu},\ and\ \citenamefont {Moshchalkov}}]{Zhang:2016_PRApp}%
		\BibitemOpen
		\bibfield  {author} {\bibinfo {author} {\bibfnamefont {G.}~\bibnamefont
				{Zhang}}, \bibinfo {author} {\bibfnamefont {T.}~\bibnamefont {Samuely}},
			\bibinfo {author} {\bibfnamefont {J.}~\bibnamefont {Ka\ifmmode \check{c}\else
					\v{c}\fi{}mar\ifmmode~\check{c}\else \v{c}\fi{}\'{\i}k}}, \bibinfo {author}
			{\bibfnamefont {E.~A.}\ \bibnamefont {Ekimov}}, \bibinfo {author}
			{\bibfnamefont {J.}~\bibnamefont {Li}}, \bibinfo {author} {\bibfnamefont
				{J.}~\bibnamefont {Vanacken}}, \bibinfo {author} {\bibfnamefont
				{P.}~\bibnamefont {Szab\'o}}, \bibinfo {author} {\bibfnamefont
				{J.}~\bibnamefont {Huang}}, \bibinfo {author} {\bibfnamefont {P.~J.}\
				\bibnamefont {Pereira}}, \bibinfo {author} {\bibfnamefont {D.}~\bibnamefont
				{Cerbu}},\ and\ \bibinfo {author} {\bibfnamefont {V.~V.}\ \bibnamefont
				{Moshchalkov}},\ }\href {https://doi.org/10.1103/PhysRevApplied.6.064011}
		{\bibfield  {journal} {\bibinfo  {journal} {Phys. Rev. Applied}\ }\textbf
			{\bibinfo {volume} {6}},\ \bibinfo {pages} {064011} (\bibinfo {year}
			{2016})}\BibitemShut {NoStop}%
		\bibitem [{\citenamefont {Postolova}\ \emph {et~al.}(2017)\citenamefont
			{Postolova}, \citenamefont {Mironov}, \citenamefont {Baklanov}, \citenamefont
			{Vinokur},\ and\ \citenamefont {Baturina}}]{Postolova:2017_SciRep}%
		\BibitemOpen
		\bibfield  {author} {\bibinfo {author} {\bibfnamefont {S.~V.}\ \bibnamefont
				{Postolova}}, \bibinfo {author} {\bibfnamefont {A.~Y.}\ \bibnamefont
				{Mironov}}, \bibinfo {author} {\bibfnamefont {M.~R.}\ \bibnamefont
				{Baklanov}}, \bibinfo {author} {\bibfnamefont {V.~M.}\ \bibnamefont
				{Vinokur}},\ and\ \bibinfo {author} {\bibfnamefont {T.~I.}\ \bibnamefont
				{Baturina}},\ }\href {https://doi.org/10.1038/s41598-017-01753-w} {\bibfield
			{journal} {\bibinfo  {journal} {Sci. Rep.}\ }\textbf {\bibinfo {volume}
				{7}},\ \bibinfo {pages} {1718} (\bibinfo {year} {2017})}\BibitemShut
		{NoStop}%
		\bibitem [{\citenamefont {Sac\'ep\'e}\ \emph {et~al.}(2015)\citenamefont
			{Sac\'ep\'e}, \citenamefont {Seidemann}, \citenamefont {Ovadia},
			\citenamefont {Tamir}, \citenamefont {Shahar}, \citenamefont {Chapelier},
			\citenamefont {Strunk},\ and\ \citenamefont {Piot}}]{Sacepe:2015_PRB}%
		\BibitemOpen
		\bibfield  {author} {\bibinfo {author} {\bibfnamefont {B.}~\bibnamefont
				{Sac\'ep\'e}}, \bibinfo {author} {\bibfnamefont {J.}~\bibnamefont
				{Seidemann}}, \bibinfo {author} {\bibfnamefont {M.}~\bibnamefont {Ovadia}},
			\bibinfo {author} {\bibfnamefont {I.}~\bibnamefont {Tamir}}, \bibinfo
			{author} {\bibfnamefont {D.}~\bibnamefont {Shahar}}, \bibinfo {author}
			{\bibfnamefont {C.}~\bibnamefont {Chapelier}}, \bibinfo {author}
			{\bibfnamefont {C.}~\bibnamefont {Strunk}},\ and\ \bibinfo {author}
			{\bibfnamefont {B.~A.}\ \bibnamefont {Piot}},\ }\href
		{https://doi.org/10.1103/PhysRevB.91.220508} {\bibfield  {journal} {\bibinfo
				{journal} {Phys. Rev. B}\ }\textbf {\bibinfo {volume} {91}},\ \bibinfo
			{pages} {220508} (\bibinfo {year} {2015})}\BibitemShut {NoStop}%
		\bibitem [{\citenamefont {Peng}\ \emph {et~al.}(2013)\citenamefont {Peng},
			\citenamefont {Meng}, \citenamefont {Mou}, \citenamefont {He}, \citenamefont
			{Zhao}, \citenamefont {Wu}, \citenamefont {Liu}, \citenamefont {Dong},
			\citenamefont {He}, \citenamefont {Zhang}, \citenamefont {Wang},
			\citenamefont {Peng}, \citenamefont {Wang}, \citenamefont {Zhang},
			\citenamefont {Yang}, \citenamefont {Chen}, \citenamefont {Xu}, \citenamefont
			{Lee},\ and\ \citenamefont {Zhou}}]{Peng:2013_NatCom}%
		\BibitemOpen
		\bibfield  {author} {\bibinfo {author} {\bibfnamefont {Y.}~\bibnamefont
				{Peng}}, \bibinfo {author} {\bibfnamefont {J.}~\bibnamefont {Meng}}, \bibinfo
			{author} {\bibfnamefont {D.}~\bibnamefont {Mou}}, \bibinfo {author}
			{\bibfnamefont {J.}~\bibnamefont {He}}, \bibinfo {author} {\bibfnamefont
				{L.}~\bibnamefont {Zhao}}, \bibinfo {author} {\bibfnamefont {Y.}~\bibnamefont
				{Wu}}, \bibinfo {author} {\bibfnamefont {G.}~\bibnamefont {Liu}}, \bibinfo
			{author} {\bibfnamefont {X.}~\bibnamefont {Dong}}, \bibinfo {author}
			{\bibfnamefont {S.}~\bibnamefont {He}}, \bibinfo {author} {\bibfnamefont
				{J.}~\bibnamefont {Zhang}}, \bibinfo {author} {\bibfnamefont
				{X.}~\bibnamefont {Wang}}, \bibinfo {author} {\bibfnamefont {Q.}~\bibnamefont
				{Peng}}, \bibinfo {author} {\bibfnamefont {Z.}~\bibnamefont {Wang}}, \bibinfo
			{author} {\bibfnamefont {S.}~\bibnamefont {Zhang}}, \bibinfo {author}
			{\bibfnamefont {F.}~\bibnamefont {Yang}}, \bibinfo {author} {\bibfnamefont
				{C.}~\bibnamefont {Chen}}, \bibinfo {author} {\bibfnamefont {Z.}~\bibnamefont
				{Xu}}, \bibinfo {author} {\bibfnamefont {T.~K.}\ \bibnamefont {Lee}},\ and\
			\bibinfo {author} {\bibfnamefont {X.~J.}\ \bibnamefont {Zhou}},\ }\href
		{https://doi.org/10.1038/ncomms3459} {\bibfield  {journal} {\bibinfo
				{journal} {Nat. Commun.}\ }\textbf {\bibinfo {volume} {4}},\ \bibinfo {pages}
			{2459} (\bibinfo {year} {2013})}\BibitemShut {NoStop}%
		\bibitem [{\citenamefont {Takagi}\ \emph {et~al.}(1992)\citenamefont {Takagi},
			\citenamefont {Batlogg}, \citenamefont {Kao}, \citenamefont {Kwo},
			\citenamefont {Cava}, \citenamefont {Krajewski},\ and\ \citenamefont
			{Peck}}]{Tagaki:1992_PRL}%
		\BibitemOpen
		\bibfield  {author} {\bibinfo {author} {\bibfnamefont {H.}~\bibnamefont
				{Takagi}}, \bibinfo {author} {\bibfnamefont {B.}~\bibnamefont {Batlogg}},
			\bibinfo {author} {\bibfnamefont {H.~L.}\ \bibnamefont {Kao}}, \bibinfo
			{author} {\bibfnamefont {J.}~\bibnamefont {Kwo}}, \bibinfo {author}
			{\bibfnamefont {R.~J.}\ \bibnamefont {Cava}}, \bibinfo {author}
			{\bibfnamefont {J.~J.}\ \bibnamefont {Krajewski}},\ and\ \bibinfo {author}
			{\bibfnamefont {W.~F.}\ \bibnamefont {Peck}},\ }\href
		{https://doi.org/10.1103/PhysRevLett.69.2975} {\bibfield  {journal} {\bibinfo
				{journal} {Phys. Rev. Lett.}\ }\textbf {\bibinfo {volume} {69}},\ \bibinfo
			{pages} {2975} (\bibinfo {year} {1992})}\BibitemShut {NoStop}%
		\bibitem [{\citenamefont {Ono}\ \emph {et~al.}(2000)\citenamefont {Ono},
			\citenamefont {Ando}, \citenamefont {Murayama}, \citenamefont {Balakirev},
			\citenamefont {Betts},\ and\ \citenamefont {Boebinger}}]{Ono:2000_PRL}%
		\BibitemOpen
		\bibfield  {author} {\bibinfo {author} {\bibfnamefont {S.}~\bibnamefont
				{Ono}}, \bibinfo {author} {\bibfnamefont {Y.}~\bibnamefont {Ando}}, \bibinfo
			{author} {\bibfnamefont {T.}~\bibnamefont {Murayama}}, \bibinfo {author}
			{\bibfnamefont {F.~F.}\ \bibnamefont {Balakirev}}, \bibinfo {author}
			{\bibfnamefont {J.~B.}\ \bibnamefont {Betts}},\ and\ \bibinfo {author}
			{\bibfnamefont {G.~S.}\ \bibnamefont {Boebinger}},\ }\href
		{https://doi.org/10.1103/PhysRevLett.85.638} {\bibfield  {journal} {\bibinfo
				{journal} {Phys. Rev. Lett.}\ }\textbf {\bibinfo {volume} {85}},\ \bibinfo
			{pages} {638} (\bibinfo {year} {2000})}\BibitemShut {NoStop}%
		\bibitem [{\citenamefont {Semba}\ and\ \citenamefont
			{Matsuda}(2001)}]{Semba:2001_PRL}%
		\BibitemOpen
		\bibfield  {author} {\bibinfo {author} {\bibfnamefont {K.}~\bibnamefont
				{Semba}}\ and\ \bibinfo {author} {\bibfnamefont {A.}~\bibnamefont
				{Matsuda}},\ }\href {https://doi.org/10.1103/PhysRevLett.86.496} {\bibfield
			{journal} {\bibinfo  {journal} {Phys. Rev. Lett.}\ }\textbf {\bibinfo
				{volume} {86}},\ \bibinfo {pages} {496} (\bibinfo {year} {2001})}\BibitemShut
		{NoStop}%
		\bibitem [{\citenamefont {Komiya}\ \emph {et~al.}(2005)\citenamefont {Komiya},
			\citenamefont {Chen}, \citenamefont {Zhang},\ and\ \citenamefont
			{Ando}}]{Komiya:2005_PRL}%
		\BibitemOpen
		\bibfield  {author} {\bibinfo {author} {\bibfnamefont {S.}~\bibnamefont
				{Komiya}}, \bibinfo {author} {\bibfnamefont {H.-D.}\ \bibnamefont {Chen}},
			\bibinfo {author} {\bibfnamefont {S.-C.}\ \bibnamefont {Zhang}},\ and\
			\bibinfo {author} {\bibfnamefont {Y.}~\bibnamefont {Ando}},\ }\href
		{https://doi.org/10.1103/PhysRevLett.94.207004} {\bibfield  {journal}
			{\bibinfo  {journal} {Phys. Rev. Lett.}\ }\textbf {\bibinfo {volume} {94}},\
			\bibinfo {pages} {207004} (\bibinfo {year} {2005})}\BibitemShut {NoStop}%
		\bibitem [{\citenamefont {Oh}\ \emph {et~al.}(2006)\citenamefont {Oh},
			\citenamefont {Crane}, \citenamefont {Van~Harlingen},\ and\ \citenamefont
			{Eckstein}}]{Oh:2006_PRL}%
		\BibitemOpen
		\bibfield  {author} {\bibinfo {author} {\bibfnamefont {S.}~\bibnamefont
				{Oh}}, \bibinfo {author} {\bibfnamefont {T.~A.}\ \bibnamefont {Crane}},
			\bibinfo {author} {\bibfnamefont {D.~J.}\ \bibnamefont {Van~Harlingen}},\
			and\ \bibinfo {author} {\bibfnamefont {J.~N.}\ \bibnamefont {Eckstein}},\
		}\href {https://doi.org/10.1103/PhysRevLett.96.107003} {\bibfield  {journal}
			{\bibinfo  {journal} {Phys. Rev. Lett.}\ }\textbf {\bibinfo {volume} {96}},\
			\bibinfo {pages} {107003} (\bibinfo {year} {2006})}\BibitemShut {NoStop}%
		\bibitem [{\citenamefont {Moshchalkov}\ \emph {et~al.}(2001)\citenamefont
			{Moshchalkov}, \citenamefont {Vanacken},\ and\ \citenamefont
			{Trappeniers}}]{Moshchalkov:2001_PRB}%
		\BibitemOpen
		\bibfield  {author} {\bibinfo {author} {\bibfnamefont {V.~V.}\ \bibnamefont
				{Moshchalkov}}, \bibinfo {author} {\bibfnamefont {J.}~\bibnamefont
				{Vanacken}},\ and\ \bibinfo {author} {\bibfnamefont {L.}~\bibnamefont
				{Trappeniers}},\ }\href {https://doi.org/10.1103/PhysRevB.64.214504}
		{\bibfield  {journal} {\bibinfo  {journal} {Phys. Rev. B}\ }\textbf {\bibinfo
				{volume} {64}},\ \bibinfo {pages} {214504} (\bibinfo {year}
			{2001})}\BibitemShut {NoStop}%
		\bibitem [{\citenamefont {Daou}\ \emph {et~al.}(2009)\citenamefont {Daou},
			\citenamefont {Doiron-Leyraud}, \citenamefont {LeBoeuf}, \citenamefont {Li},
			\citenamefont {Lalibert{\'e}}, \citenamefont {Cyr-Choini{\`e}re},
			\citenamefont {Jo}, \citenamefont {Balicas}, \citenamefont {Yan},
			\citenamefont {Zhou}, \citenamefont {Goodenough},\ and\ \citenamefont
			{Taillefer}}]{Daou:2009_NatPhys}%
		\BibitemOpen
		\bibfield  {author} {\bibinfo {author} {\bibfnamefont {R.}~\bibnamefont
				{Daou}}, \bibinfo {author} {\bibfnamefont {N.}~\bibnamefont
				{Doiron-Leyraud}}, \bibinfo {author} {\bibfnamefont {D.}~\bibnamefont
				{LeBoeuf}}, \bibinfo {author} {\bibfnamefont {S.~Y.}\ \bibnamefont {Li}},
			\bibinfo {author} {\bibfnamefont {F.}~\bibnamefont {Lalibert{\'e}}}, \bibinfo
			{author} {\bibfnamefont {O.}~\bibnamefont {Cyr-Choini{\`e}re}}, \bibinfo
			{author} {\bibfnamefont {Y.~J.}\ \bibnamefont {Jo}}, \bibinfo {author}
			{\bibfnamefont {L.}~\bibnamefont {Balicas}}, \bibinfo {author} {\bibfnamefont
				{J.-Q.}\ \bibnamefont {Yan}}, \bibinfo {author} {\bibfnamefont {J.-S.}\
				\bibnamefont {Zhou}}, \bibinfo {author} {\bibfnamefont {J.~B.}\ \bibnamefont
				{Goodenough}},\ and\ \bibinfo {author} {\bibfnamefont {L.}~\bibnamefont
				{Taillefer}},\ }\href {https://doi.org/10.1038/nphys1109} {\bibfield
			{journal} {\bibinfo  {journal} {Nat. Phys.}\ }\textbf {\bibinfo {volume}
				{5}},\ \bibinfo {pages} {31} (\bibinfo {year} {2009})}\BibitemShut {NoStop}%
		\bibitem [{\citenamefont {Zaken}\ and\ \citenamefont
			{Rosenbaum}(1994)}]{Zaken:1994_JPCM}%
		\BibitemOpen
		\bibfield  {author} {\bibinfo {author} {\bibfnamefont {E.}~\bibnamefont
				{Zaken}}\ and\ \bibinfo {author} {\bibfnamefont {R.}~\bibnamefont
				{Rosenbaum}},\ }\href {https://doi.org/10.1088/0953-8984/6/46/016} {\bibfield
			{journal} {\bibinfo  {journal} {J. Phys. Condens. Matter}\ }\textbf
			{\bibinfo {volume} {6}},\ \bibinfo {pages} {9981} (\bibinfo {year}
			{1994})}\BibitemShut {NoStop}%
		\bibitem [{\citenamefont {Kalenkov}\ \emph {et~al.}(2012)\citenamefont
			{Kalenkov}, \citenamefont {Zaikin},\ and\ \citenamefont
			{Kuzmin}}]{Kalenkov:2012_PRL}%
		\BibitemOpen
		\bibfield  {author} {\bibinfo {author} {\bibfnamefont {M.~S.}\ \bibnamefont
				{Kalenkov}}, \bibinfo {author} {\bibfnamefont {A.~D.}\ \bibnamefont
				{Zaikin}},\ and\ \bibinfo {author} {\bibfnamefont {L.~S.}\ \bibnamefont
				{Kuzmin}},\ }\href {https://doi.org/10.1103/PhysRevLett.109.147004}
		{\bibfield  {journal} {\bibinfo  {journal} {Phys. Rev. Lett.}\ }\textbf
			{\bibinfo {volume} {109}},\ \bibinfo {pages} {147004} (\bibinfo {year}
			{2012})}\BibitemShut {NoStop}%
		\bibitem [{\citenamefont {Shelly~Connor}\ \emph {et~al.}(2016)\citenamefont
			{Shelly~Connor}, \citenamefont {Matrozova~Ekaterina},\ and\ \citenamefont
			{Petrashov~Victor}}]{Shelley:2021_SciAdv}%
		\BibitemOpen
		\bibfield  {author} {\bibinfo {author} {\bibfnamefont {D.}~\bibnamefont
				{Shelly~Connor}}, \bibinfo {author} {\bibfnamefont {A.}~\bibnamefont
				{Matrozova~Ekaterina}},\ and\ \bibinfo {author} {\bibfnamefont
				{T.}~\bibnamefont {Petrashov~Victor}},\ }\href
		{https://doi.org/10.1126/sciadv.1501250} {\bibfield  {journal} {\bibinfo
				{journal} {Sci. Adv.}\ }\textbf {\bibinfo {volume} {2}},\ \bibinfo {pages}
			{e1501250} (\bibinfo {year} {2016})}\BibitemShut {NoStop}%
	\end{thebibliography}


\end{document}